\documentstyle[11pt,aaspp4,tighten]{article}

\lefthead{Hillenbrand {\it et al.}}
\righthead{Orion Nebula Cluster}

\begin{document}


\title{Constraints on the Stellar/Sub-stellar Mass Function 
    \\ in the Inner Orion Nebula Cluster}
\author{Lynne A. Hillenbrand\altaffilmark{1} 
	and
        John M. Carpenter\altaffilmark{1}}
\affil{California Institute of Technology}
\authoraddr{Dept. of Astronomy; MS 105-24; Pasadena, CA 91125}
\authoremail{lah@astro.caltech.edu}
\altaffiltext{1}{Visiting Astronomer, W.M. Keck Observatory, 
which is operated as a scientific partnership among the
California Institute of Technology, the University of California 
and the National Aeronautics and Space Administration. 
The Observatory was made possible by the generous financial support 
of the W.M. Keck Foundation.}


\newcommand\khk{K--(H-K)\ }

\begin{abstract}
We present the results of a 0.5-0.9$"$ FWHM imaging survey 
at K (2.2$\mu$m) and H (1.6$\mu$m) covering $\sim$5.1'$\times$5.1'
centered on $\theta^1C$ Ori, the most massive star in the Orion Nebula Cluster
(ONC).
At the age and distance of this cluster, and in the absence of extinction,
the hydrogen burning limit (0.08 M$_\odot$) occurs at
K$\approx$ 13.5 mag while an object of mass 0.02 M$_\odot$
has K$\approx$ 16.2 mag.  Our photometry is complete
for source detection at the 7$\sigma$ level to K$\approx$17.5 mag and 
thus is sensitive to objects as low-mass as 0.02$M_\odot$ seen through
visual extinction values as high as 10 magnitudes.  We use the observed 
magnitudes, colors, and star counts to constrain the shape of the inner ONC 
stellar mass function across the hydrogen burning limit.  
After determining the stellar age and near-infrared excess properties
of the optically visible stars in this same inner ONC region,
we present a new technique that incorporates these distributions when
extracting the mass function from the observed density of
stars in the K--(H-K) diagram.
We find that our data are inconsistent with a mass function that rises 
across the stellar/sub-stellar boundary.  Instead, we find that the most likely
form of the inner ONC mass function is one that rises to a peak around 
0.15 M$_\odot$, and then declines across the hydrogen-burning limit 
with slope N(log M) $\propto$ M$^{0.57}$.
We emphasize that our conclusions apply to the inner 0.71 pc x 0.71 pc 
of the ONC only; they may not apply to the ONC as a whole where some evidence 
for general mass segregation has been found.
\end{abstract}


\section{Introduction}

The Orion Nebula is one of the most famous objects in the sky, and has been 
the target of innumerable astronomical observations at virtually all wavelengths
over the past 100 years.  Yet it is only within the past few years that we have
begun to discover the extent of the young stellar population just emerging 
from the ambient molecular cloud, and to characterize its nature. Work by
Herbig \& Terndrup (1986), Prosser et al. (1994) and Hillenbrand (1997) 
has established that the mean age of stars projected within $\sim$2 pc 
of the massive Trapezium stars is $<$1 Myr.  
The mass distribution derived for $\sim$1000 ONC stars located on a theoretical
HR diagram by Hillenbrand (1997) rises to $\sim0.2~M_\odot$ and shows
some evidence for flattening or turning over towards lower masses (see, however,
the reinterpretation of these data in Figure~\ref{fig:optical.newvsold}
of the current paper using updated tracks/isochrones and 
updated transformations from observational to theoretical quantities).
Our former study was complete to just above the hydrogren burning limit and
did not constrain the mass function across
the stellar/sub-stellar boundary into the brown dwarf regime. 
Existence of brown dwarfs in the ONC has been discussed previously by
McCaughrean et al. (1995).

Star forming regions like the ONC provide one of the best environments for
investigating the shape of the stellar mass function into the brown dwarf 
regime.  Unlike the case in older clusters and associations,
star-forming regions are essentially unperturbed 
by dynamical evolution that selectively remove the lowest mass objects. 
Further, contracting low-mass
pre-main sequence stars and brown dwarfs are 2-3.5 orders of magnitude more
luminous that their counterparts on the main sequence and hence can be
readily detected, especially in the near-infrared. Star forming regions are 
also less affected by field star contamination compared to older clusters due to
their small angular extent and their association with obscuring molecular 
material.  The ONC cluster in particular 
affords several distinct advantages compared to any other young stellar cluster
for measuring the initial mass function. 
First, since it is located at high galactic latitude toward the outer Galaxy, 
contamination from field stars is minimized. Further, the winds and ionization 
from the central OB stars have dispersed much of the surrounding gas and dust, 
drastically reducing the extinction to the cluster members. A high-column
density of obscuring molecular material does remain intact behind the stellar
cluster. Foremost, however, 
as the nearest massive star-forming region to the Sun and the most populous 
young cluster within at least 2 kpc, the ONC is the one region where one can
assemble a statistically robust assessment of the mass distribution well
into the brown dwarf regime.

In this contribution we investigate whether the distribution of stars in the 
\khk color-magnitude diagram for the ONC is consistent
with a mass function that rises across the stellar/sub-stellar boundary and into
the brown dwarf regime, or if the data demand that the mass function 
turns over.  After describing the observations, image analysis, construction of
the point source list, and extraction of photometry, we present a new
approach for constraining the stellar/sub-stellar mass function.  
We  consider that the location of a particular star in the \khk diagram 
depends on four parameters:  stellar mass, stellar age, presence and properties 
of a circumstellar disk, and extinction.  De-reddening the stars along a known
reddening vector in the K-(H-K) diagram enables us to compute the probability 
that a star could be of a certain mass given the distributions in age and 
near-infrared excess that characteristize the ONC cluster. 
Summing of these individual mass
probability distributions yields the mass function for the entire cluster. 
We believe that our technique produces the most rigorously derived constraint 
yet from photometry alone on the inner ONC initial mass function.

\section{Observations}

Images were obtained on 8, 9, 10 February, 1999
using NIRC (Matthews \& Soifer 1994) mounted on the Keck I 10-m telescope.
Data were taken in H-band (1.50-1.82$\mu$m)
on the first night, K- (2.00-2.43$\mu$m) and H-bands on the second night, 
and Z-band (0.95-1.11$\mu$m) on the third night.
The field of view of a NIRC frame is 38$"$ x 38$"$ at 0.152$"$/pixel.
For each filter, the observations consisted of a 15 x 15 grid of such frames
aligned with the equatorial coordinate system to produce a 5.1$'$ x 5.1$'$
mosaic. Adjacent rows and columns in the grid were spaced by one-half of the
array, so that any one pixel within the final mosaic was nominally observed
on four different frames (modulo minor telescope drift). The integration
times per frame were 0.5 sec with 50 co-adds (25 seconds total) at H- and
K-bands, and 2 sec with 20 coadds (40 seconds total) at Z-band. Such short
integrations per frame were necessitated by the large number density of
relatively bright (K $<$ 12 mag) stars in the region whose saturation effects 
we wished to minimize.

The observing sequence for each declination row in the grid was to center the
array on a previously chosen setup star, offset to the beginning of a row,
scan in right ascension across the row, then offset to and five point dither on an
off-field sky position, return to the setup star, and repeat.  Local sky was 
measured after every row (at 10-12 minute spacings in time) from a location
$\sim$15' northwest of the ONC which we had determined via examination 
of the 2MASS Image Atlas to be free of nebulosity and relatively free 
of infrared point sources.  We chose to include in our sky field 
a star of magnitude K$\approx$14 mag in order to monitor the atmospheric
extinction with airmass locally (see, however, the results in Appendix A).  
In addition, we observed absolute photometric
standards from Persson et al. (1998).  Sky conditions on all three nights
were photometric, with standard star solutions matching
nominal NIRC zero points and nominal Mauna Kea extinction curves with airmass.
Flat-field, bias, and linearity calibration data were also obtained.

\section{Image Processing}

The NIRC images were processed in IRAF first by determining the detector gain 
and readnoise from raw flat-field and bias frames, and establishing the 
linearity from a series of exposures taken with different integration times
of the tertiary mirror cover.  The latter tests indicated NIRC is linear between within 
0.5\% up to 90\% of the full-well depth.
Next, a median-filtered, normalized flat-field image was constructed for each 
filter from a series of 10 bias-subtracted dome flats.  Bad pixel masks were made
from the flat-fields by identifying all pixels more than 12\% above or 15\%
below the mean value.  Approximately 1.5\% of the pixels were flagged as bad,
with most of these located in a 1 pixel border around the edge of the array.
Sky images were constructed for each of the 15 rows in our mosaic 
(in each filter) using the 5 dithered off-field sky frames, 
after dark-subtracting, median-filtering, and bad pixel exclusion.  

Each of the on-field ONC data images 
was then sky- and dark-subtracted, and flat-divided.  Next, on a frame-by-frame
basis we identified and interpolated over intermittently appearing
``warm" pixels.  These pixels had values 20-50\% above the local background 
and thus would affect our averaging of overlapping pixels in the final mosaic.
The ``warm" pixels numbered between $\sim$5-30 per frame with
no discernible pattern and appeared as ``warm" for a sequence of $\sim$3-10 
frames before returning to normal values.  

Next, we corrected for a feature of NIRC known as ``bleeding''
(Liu \& Graham 1997). 
The readout electronics of NIRC are such that bright stars exhibit
an exponentially decaying trail to the right which wraps around the right edge
of the array and continues from the left, one row higher.
A similar effect trailing downwards and wrapping around to the top of the array
is associated with the brightest of stars.  This ``bleeding'' behavior must 
be present at a visually imperceptible level for every star, 
and must be in the background counts as well. It can be thought of as part of 
the point-spread-function.  But we need to model and correct for it 
since ``bleeding'' from the brighter stars can extend over other
sources in the field and, furthermore, will adversely affect our frame-to-frame
flux adjustments in the mosaicing process and our ability to model the
point-spread-function using standard methods.  We have used the empirical 
solution developed by Liu \& Graham of subtracting from every pixel 
in the array, the exponentially decaying contribution of every pixel 
further back in the NIRC readout scheme. The coefficients of the correction
are such that each pixel contributes 0.25\% of its counts to the next pixel
which is read out (physically located four pixels to the right on the array)
with a pixel of, for example, 20,000 ADU contributing 51 ADU to a pixel
4 downstream and 1 ADU to a pixel 256 (1 row) downstream.
Application of this ``de-bleeding'' method does not affect the photometry,
or if it does, the level of any difference is well less than 1\%. 

Next, we corrected the frames for the effects of optical path distortion 
which amounts to $\sim$1 pixel from the center to the edge of the array.
A. Ghez generously provided to us a subroutine for this step.
Correction for image distortion improves the photometry by 0.02$\pm$0.01 mag.
The final step in the raw frame processing was to correct each image
by a multiplicative factor representing the flux adjustment from the
observed airmass to zero airmass.  At K-band this factor was
$10^{0.4 \times 0.092 \times AIRMASS}$ while at H-band it was 
$10^{0.4 \times 0.065 \times AIRMASS}$.

Custom C programs were written to co-add the 225 images in each band into
a mosaic in order to detect faint point sources and to improve the 
signal-to-noise of the photometry. The mosaics were constructed by determining
the relative positional and sky offsets between the overlapping frames, 
and making the appropriate shifts in order to tile them together. 

Positional offsets were established using stars identified 
in the overlap regions of neighboring frames. To reduce the random-walk errors 
in stitching the images together, the relative offsets of all 225 frames per
band were solved simultaneously using a linear least-squares fit that 
minimized the position residuals for all stellar matches in all the 
overlap regions.
Histograms of the resulting positional residuals at K band have a 
1$\sigma$ standard deviation of 0.017$''$ in right ascension and 0.028$''$ in 
declination, and 0.022$''$ and 0.026$''$, respectively, for H band.

\begin{figure}
\vskip -0.5truein
\plotfiddle{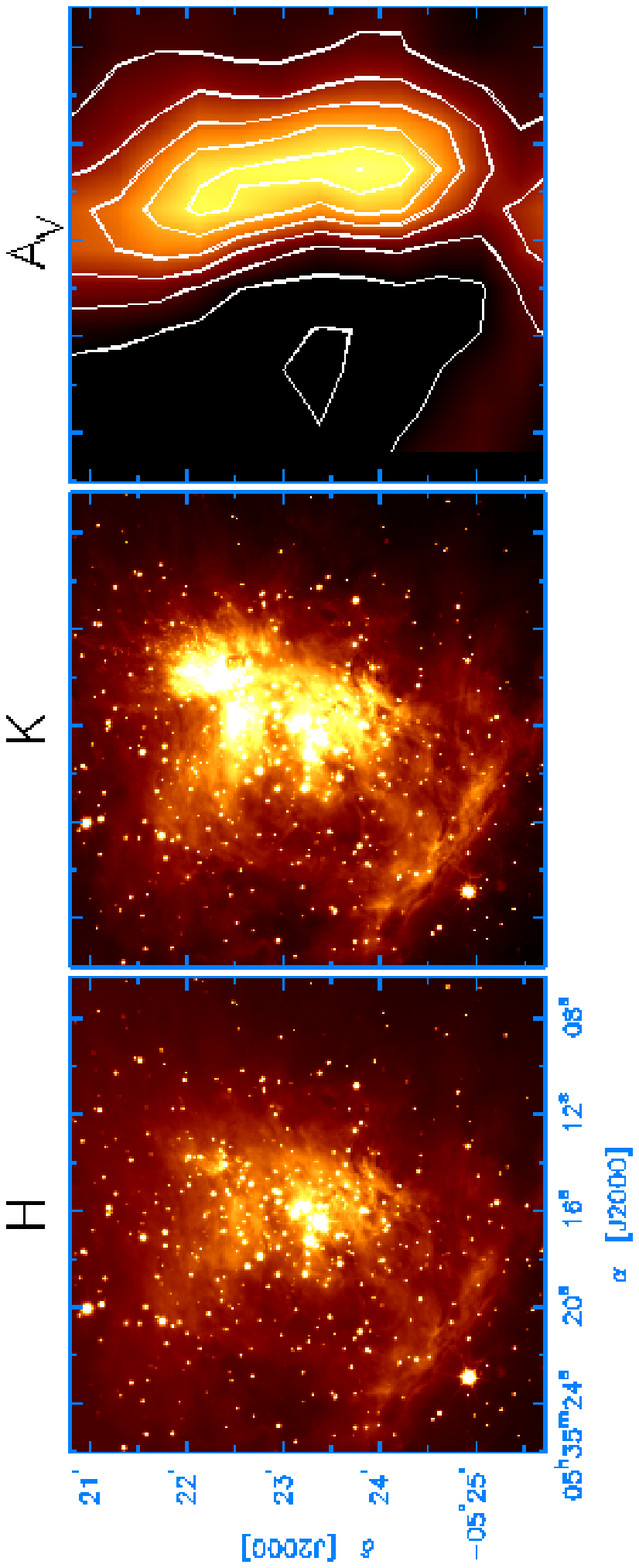}{5.in}{-90}{75}{75}{-260}{400}
\figcaption{ Images of our H and K-band mosaics from Keck/NIRC along with 
an extinction map derived
from the molecular column density data of Goldsmith, Bergin, \& Lis (1997).
The pixel size of the infrared mosaics is 0.15$"$ and the angular
resolution of the extinction map is 50$"$. Contours in the
extinction map begin at A$_V$ = 5 mag and are spaced at 
$\Delta$A$_V$ = 10 mag intervals.
\label{fig:mosaics}
}
\end{figure}

\begin{figure}
\epsscale{1.1}
\plotone{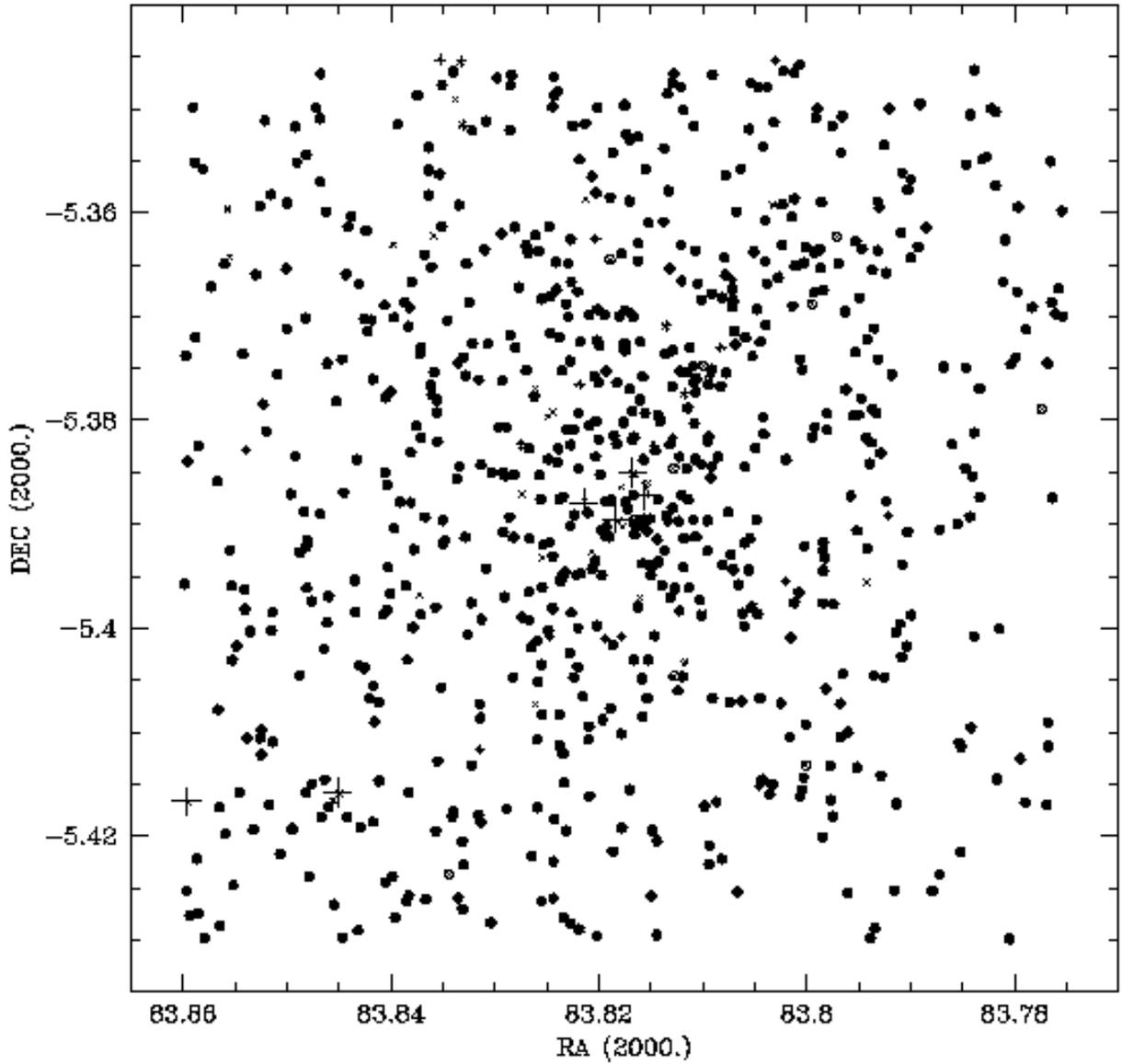}
\vskip -1.5truein
\figcaption{Spatial distribution of ONC stars within our 
NIRC mosaics. The $x$'s indicate stars whose photometry we
could not derive, $x$'s surrounded by open circles indicate stars with
photometry at K but not H, $\ast$'s indicate stars with photometry at
H but not K, and filled circles indicate stars with photometry at both K and
H. Large $+$ signs indicate the optically brightest stars, for orientation.
\label{fig:space}
}
\epsscale{1.00}
\end{figure}

As the images were placed in the mosaic, the sky background was adjusted 
by an additive constant to match the background in the surrounding frames. The 
relative intensity offsets were determined by fitting a gaussian to the 
difference in image intensity in the overlap region between neighboring frames.
As with the positional offsets, the sky offsets for all frames were determined
simultaneously using a linear least-squares fit to all the overlap regions.
We were unable to obtain an acceptable solution over the entire mosaic
for the sky offsets at Z-band.  Thus our Z-band photometry is derived
from the individual frames and is not as deep as it would be if derived
from a co-added mosaic.  The Z-band data and supplemental calibration
information are presented in a separate paper.
The H- and K-band mosaics are shown in Figure~\ref{fig:mosaics} along with 
an extinction map which is described in section 5.2.  Figure~\ref{fig:space}
shows the spatial distribution of stars in our sample, whose identification
and photometry we describe next.  Table~\ref{tab:data} contains the
coordinates and HK photometry.

\section{Mosaic Analysis}

\subsection{Identification of Point Sources}


Point sources were identified on the K-band mosaic
using DAOFIND in IRAF with a 7$\sigma$ threshold.
The initial source list was hand-edited to remove nebular knots,
multiple listings of bright stars, diffraction spikes, edge effects, etc. 
We then examined contour plots of each point source to look for
extended or double-peaked structure, and added any newly found sources.
The final source list consists of 778 
stellar point sources over the 5.1$'$x5.1$'$ field.

\subsection{Aperture and Point-Spread Function Fitting Photometry}

Aperture photometry was derived using PHOT with a 6 pixel radius aperture 
and a sky annulus extending radially from 7-12 pixels; contribution 
from the sky was determined from the mode of these values.  The small 
aperture and the close sky annulus were necessitated mainly by the 
spatially variable background from the Orion Nebula, and to a lesser 
extent by the high source density.  
Aperture corrections were needed in order to correct from the
6 pixel radius used to measure the data to the 20 pixel radius 
used to measure the standard stars.
The size of the aperture correction is directly related to the size
(e.g. full-width-half-maximum; FWHM) of the point source.  In our image
mosaics, however, the point-spread-function (PSF) undergoes 
large and non-systematic spatial variations due to random wandering 
in time of the seeing compared to the 0.15$"$ platescale, and to
a systematic gradient in airmass.  We derived an empirical calibration
between the aperture correction and the image FWHM, as follows.

First, recall that each stellar image in our final mosaic is synthesized 
from several (ideally four) separate observations of the star.  Each of these
observations may have a different PSF size. We verified that the PSFs of 
point sources in the final mosaic indeed have the mean value of the PSFs 
and the mean value of photometry through a fixed aperture size, 
characterizing the individual images from which they were created.  
In order to determine the appropriate aperture corrections for photometry 
from the mosaiced data, we therefore need only measure each of the PSFs 
in the final mosaic and apply a correlation between PSF and 
aperture corrections.


We measured the size of each stellar PSF using
a variety of IRAF tasks -- IMEXAMINE, RADPROF, and FITSPSF. 
From the 40 most isolated K $<$ 13 mag stars
in the mosaic we found the tightest correlation (error in slope $<$ 0.01 mag)
to be between the aperture correction and the ``enclosed'' gaussian fit FWHM
of the IMEXAMINE task.  In cases where this primary FWHM -- aperture correction
correlation could not be applied (e.g. failure of the gaussian fit to
converge due to crowding and/or high nebulosity) we used secondary 
correlations.  The FWHM values of $>$90\% of the stars are between 3-5 pixels at 
K-band and between 4.5-6.5 pixels at H-band.  The full range of the aperture corrections 
at K is from $\sim$ -0.10 mag to -0.60 mag with a mean value of -0.350 mag
and at H is from $\sim$ -0.25 to -0.55 mag with a mean value of -0.325 mag.
Variable aperture corrections are less of a problem at H compared to K
since the overall size of the stellar images is larger 
which acts to decrease the percentage of PSF change as the 
seeing and airmass vary.  Despite the complicated nature of our process
for applying aperture corrections, we believe it is the proper one based on 
significant improvement in correlations between our NIRC photometry 
and previous photometry (described below). 
We did attempt to correlate aperture corrections 
with a measure of the difference in width between each stellar image 
and a PSF constructed from the data itself (the ``sharpness'' parameter 
of the PEAK task). But the correlation was too loose to be useful.

The point-spread-function was constructed using the PSF task and 30 stars 
distributed over the outer, less crowded, regions of our mosaics and having 
FWHM values distributed like the data as a whole.  
Accurate characterization of the noise in the mosaiced images was done
so as to achieve the best possible fits of the point-spread-function to the 
stellar sources.  In practice, this means adding a constant to the mosaics
such that their standard deviation equals the sky counts plus the square
of the effective read noise.  
A moffat function with $\beta$=1.5 gave the best residuals among the moffat, 
lorentzian, gaussian, and penny functions.  As just discussed, 
the point-spread-function varies across our image due to seeing fluctuations
and airmass changes. Because the variations are random and not smooth,
we can not model them 
in any useful way and we are forced to fit the same point-spread-function
to every star.  The constructed point-spread-function was fit using the PEAK 
task with a fitting radius equivalent to twice the average FWHM.
Both positional re-centering and sky re-calculation were permitted.
Unfortunately, the resultant photometry was strongly influenced by 
which star was chosen as the first in constructing the PSF.   

Comparisons between the aperture photometry and the point-spread-function
fitting photometry are generally poor, due to the varying PSF.  
We have decided to use in our analysis the results from aperture photometry 
calibrated as described above.
Our final photometry list was hand-edited to remove the measurements in 
cases of contamination from bright stars and/or overlapping apertures (44 stars)
or nonlinearity (36 stars). 

\subsection{Integrity of Photometry}

In this section we discuss the internal errors of our photometry and 
the comparison of our photometry to previous work.

\begin{figure}
\vskip -0.5truein
\plotone{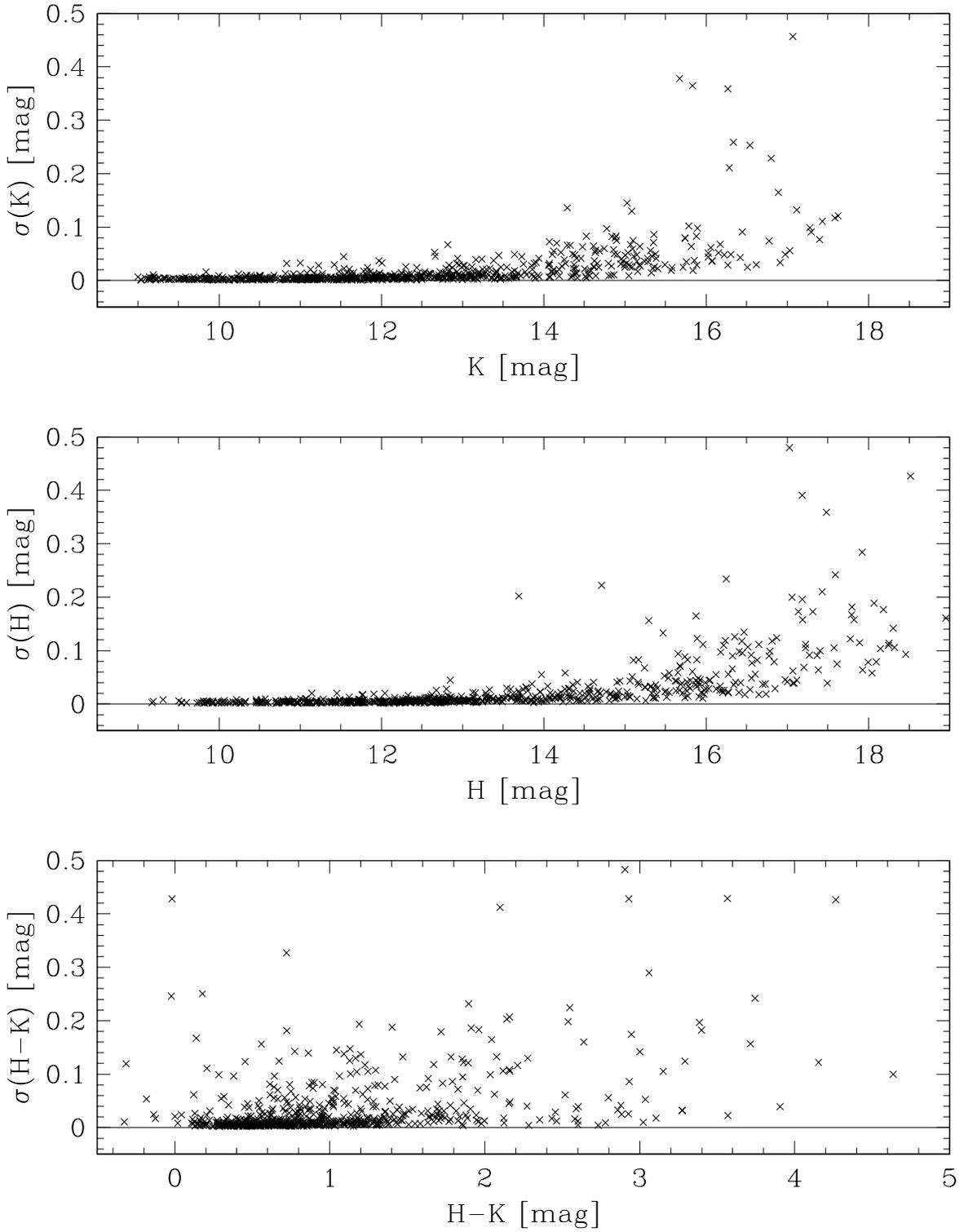}
\figcaption{Internal (IRAF) errors in photometry at H- and K-band, and in 
H-K color.
\label{fig:errvsmag}
}
\end{figure}

Figure~\ref{fig:errvsmag} shows the run of photometric error with magnitude 
and color.  These internal errors are those produced by the PHOT routine
in IRAF and simply reflect photon statistics of the source and sky 
determination in the fully processed images; 
they do not include other errors such as those in the zero point.
At K-band, 75\% of the stars in our source list have errors $<$0.02 mag, 
90\% have errors $<$0.05 mag, and 97\% have errors $<$0.1 mag. 
At H-band, 69\% of the stars in our source list have errors $<$0.02 mag, 
85\% have errors $<$0.05 mag, and 91\% have errors $<$0.1 mag. 
For the PSF fitting photometry the percentages of stars with internal errors
of the magnitudes given above were all down by 5 to 45 points, with the worst 
results at K-band.  This is due to the generally poor 
fit of any single point-spread-function to all stellar images in the mosaic
and part of our justification for rejecting the PSF photometry.

\begin{figure}
\vskip -0.6truein
\plotone{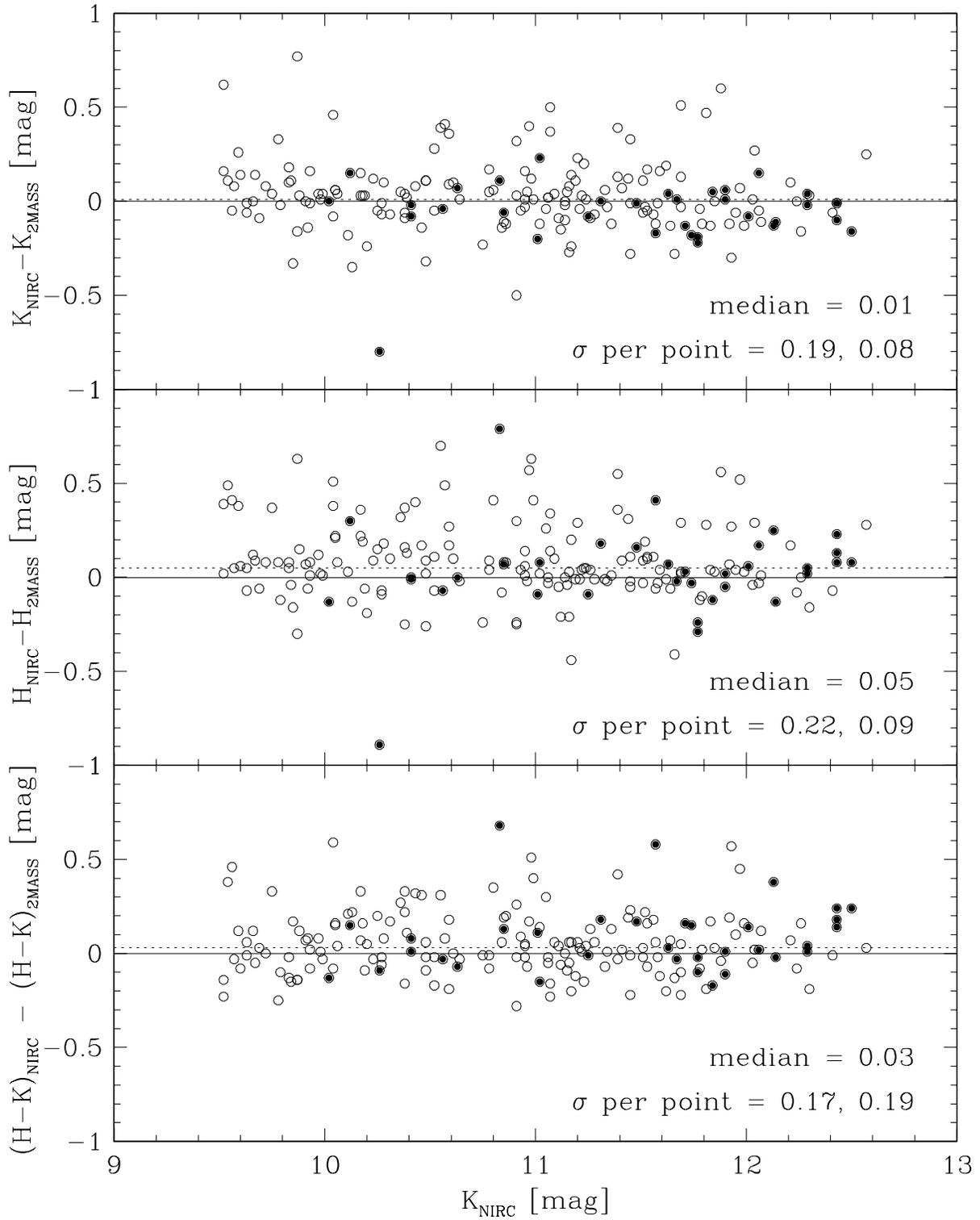}
\vskip -0.35truein
\figcaption{Comparison of NIRC and 2MASS photometry.
Open circles represent all positional matches $<$ 1" between our NIRC
sources and 2MASS sources while filled circles represent a set of relatively
bright, isolated stars (those used to derive the aperture corrections).
At K, the standard deviation per point about the mean is 0.19 mag 
for the full sample but 0.08 mag for the isolated stars.
At H, the standard deviations are 0.22 mag and 0.09 mag for the full sample
and for the isolated stars.  In H-K, the values are 0.17 mag and 0.19 mag. 
\label{fig:compare2mass}
}
\end{figure}

Figure ~\ref{fig:compare2mass}
shows the comparison of our photometry to photometry from the 
2MASS survey.  Although the scatter is large ($\sim$0.2 mag) for the full
sample of stars in common, it drops to $<$0.1 mag when only spatially
well-isolated stars are considered (filled circles).   
Many of the largest deviations ($>$1 mag) are found in crowded regions where 
our photometry is always fainter than the 2MASS photometry, presumably
due to our higher spatial resolution which permits better source separation
and better sky determination.  Nonetheless, there are also well-isolated
stars with rather large differences in the photometry.  Note, for example,
the star at K$_{NIRC}$=10.3, K$_{NIRC}$-K$_{2MASS}$=-0.8.  This is a
very bright, very well-isolated star whose photometry differs by almost 
1 mag at K between our NIRC data, the 2MASS data, our previous photometry
with SQIID/NICMASS (Hillenbrand et al. 1998), and that published 
by Ali \& DePoy (1995) and Hyland, Allen, \& Bailey (1993; as reported by
Samuel 1993).  Incidentally, this star (JW 737) is a 4.5 mag variable at 
I-band (W. Herbst, private communication). 
We have no choice but to interpret cases such as this as examples 
of real infrared photometric variability.  Variability may well be the cause 
of much of the spread along the ordinate in Figure~\ref{fig:compare2mass}.  
Indeed, we have strong evidence from Figure~\ref{fig:variablestar} 
(discussed in the Appendix) that short-term variations of order 0.1 mag
are present in at least some stars in the Orion A molecular cloud.
Comparisons between our NIRC photometry and that presented 
in Hillenbrand et al. (1998), and between 2MASS and Hillenbrand et al. (1998),
show somewhat larger scatter in the magnitudes but similar scatter in
the colors.

In summary, we believe that our photometry for bright, isolated stars
is within $<$0.1 mag of that determined by others.  Much of this scatter
may be attributed to photometric variability, although some role is probably
played by the variable PSF which plagued our photometry extraction.  We note 
that our careful attention to aperture corrections improved considerably
the correlations between our NIRC results, our previously
published data, and the 2MASS survey.  Based on the comparisons in 
Figure~\ref{fig:compare2mass}, however, for the analysis presented below 
we conservatively assume minimum K magnitude errors of $0.09/\sqrt{2}=0.06$ mag 
and minimum H-K color errors of $0.18/\sqrt{2}=0.13$ for all stars 
in our sample.  The $\sqrt{2}$ factor implies that we ascribe equal
errors to our data and to 2MASS data even though the formal 2MASS errors
are much larger than the formal NIRC errors for these relatively bright stars.

\subsection{Artificial Star Experiments}

We determined the completeness of our point source list and the completeness
of our photometry using results from extensive experimentation 
with artificial stars.  
First, fake source lists were generated by randomly distributing 300 stars 
over the $\sim$2080 $\times$ 2080 pixels$^2$ in our mosaics, 
with the caveat that no star be placed within 3 pixels of a known
star or within 30 pixels of the edge of the frame.  Next, stars with the
point-spread-function derived as described above were added to the image
at these 300 locations using ADDSTAR.  The enhanced image was then run through 
the DAOFIND, PHOT, and PEAK tasks in a manner identical to that used to
extract photometry from the unaltered data.  A separate artificial star test
was conducted for every 0.25 magnitude interval in the range 14-18.5 mag. 
Our results for {\it finding} fake stellar point sources are somewhat different
from our results for {\it photometering} fake stellar point sources.  

The DAOFIND results are that for a detection threshold of 20$\sigma$,
we can identify fake stellar point sources in our images at the 90\% completeness 
level down to K = 16.8 mag and at the 20\% completeness level down to 
K = 17 mag. For a detection threshold of 7$\sigma$, we can identify fake 
stellar point sources at the 90\% completeness level down to K = 17.7 mag  
and at the 20\% completeness level down to K = 18.2 mag. 

The DAOPHOT results are presented in Tables~\ref{tab:internalerrs} 
and ~\ref{tab:externalerrs}. We assess both the internal errors, 
the formal uncertainties in the output magnitudes, and the external 
errors, the differences between the input and the recovered magnitudes, 
as a function of magnitude and also radial position in the cluster.  The strong 
and variable nebular background in combination with extreme point source 
crowding makes these experiments somewhat more difficult to interpret
than in the usual case.  Nevertheless, we conclude based on internal error 
estimates (Table ~\ref{tab:internalerrs}) 
that at the limit of our ability to detect 90\% of the stellar point 
sources (K = 16.8 mag for the 20$\sigma$ threshold which produces 94\% of the
stars in our source list), we are able to do photometry accurate to 0.02 mag 
for 15\% of them, 0.05 mag for 68\% of them, and 0.1 mag for 85\% of them.   
For brighter stars the internal error estimates are lower.  For fainter stars, 
at the K = 17.7 mag limit 7$\sigma$ detection threshold, the photometry
is accurate to 0.02 mag for 0\% of the sources, to 0.05 mag for 19\% of them, 
and 0.1 mag for 59\% of them.   However, based on the external errors
(Table ~\ref{tab:externalerrs}), we seem perfectly capable of recovering 
the input stellar magnitudes to within a few percent down to K$\approx$17.5 mag.
The numbers listed in each column of this table are the median offsets 
between input and output magnitudes, and also the standard deviations.
Of note is that any bias 
in our photometry, when it appears at a moderately important level for stars 
K $<$ 17.5 mag, is such that the MEDIAN offset is positive, meaning that 
we seem to measure the star as slightly fainter than it really is, 
probably through over-subtraction of background.  
By contrast, the MEAN offsets are positive (meaning that we measure 
the star as being too bright), but we note that the MEAN values 
are dominated by just a few data points with special problems 
such as proximity to very bright stars; hence we prefer to quote MEDIAN
offsets. 

In conclusion, based on artificial star experiments
we adopt a conservative K = 17.5 mag as the completeness 
limit for source detection.  The 75\% completeness limits for photometry 
accurate to 10\% are K = 17.3 mag and H = 17.4 mag  
while the 50\% numbers for photometry
accurate to 10\% are K = 18.0 mag and H = 18.1 mag.  
These numbers are averaged over the 5.1$'$ x 5.1$'$ field and mask
the systematic gradient with radius caused by variable source crowding
and nebular strength.  We estimate that we have determined
the completeness to an accuracy of a few tenths of a magnitude only,
due in part to the radial gradient and in part to the spatially variable PSF.
Note that we found our ability to reproduce the magnitudes assigned to fake 
stars in the input stage varied significantly with the parameters given 
to the PHOT, PSF, and PEAK tasks in IRAF.  The parameters producing
the most accurate results in the artifical star experiments 
were those then used to extract the real source photometry as described
previously.

\subsection{Astrometry}

Our astrometry for the NIRC mosaics is referenced to the 2MASS database,
which in turn is referenced to the ACT catalog. The nominal 1$\sigma$ 2MASS position
error for bright, isolated sources is $\sim$ 0.1$"$ in each of right ascension
and declination. An edited list
of $\sim$230 stars in common between 2MASS and our Table~\ref{tab:data} was
used to derive the final astrometric solution, producing a platescale of
0.152$"$/pixel and a total r.m.s. error in the positions of 0.10$"$.


We note that the current astrometric system is shifted by about 
+1.5$"$ in right ascension and -0.3$"$ in declination compared to that presented
by us previously.  In an optical study of stars located within $\sim$20' 
of the ONC core, Hillenbrand (1997) derived astrometry using 
the HST Guide Star Catalog (Lasker et al. 1988) which is known to suffer
some inaccuracies in this region of the sky.  We have found that our previous
astrometric solution is offset to the west and to the south 
compared to other studies of the ONC (e.g. Jones \& Walker 1988;
McCaughrean \& Stauffer 1994; Prosser et al. 1994; Ali \& DePoy 1995;
O'Dell \& Wong 1996)
but by various amounts as several of these studies have their own astrometric
inaccuracies.  The Jones \& Walker positions and the McCaughrean \& Stauffer 
positions are each internally consistent, although offset from one another. 
The McCaughrean \& Stauffer astrometry matches our 2MASS-referenced astrometry.
The HST positions of Prosser et al. and of O'Dell \& Wong, however, are not 
internally consistent and suffer random excursions of about 1$"$ which were 
propogated into the Hillenbrand (1997) database.   
Likewise, the coordinates of Ali \& DePoy suffer large random errors, 
of order 1.5-2$"$. Furthermore, approximately 1/2 of the sources 
supposedly located within our survey area as listed by Ali \& DePoy 
simply do not exist in our higher resolution data; we do not discuss this catalog further.
In summary, we believe that the positions quoted in Table~\ref{tab:data}
are both internally consistent and properly referenced to the ACT 
reference frame.  


\subsection{Properties of Final Source List}

A fundamental result of this paper is a list of coordinates 
and available HK photometry for 778 stars in the inner 5.1$'$ x 5.1$'$ of the 
ONC.  A total of 687 stars have measurements at both H and K, with 647 stars
having errors $<$0.15 mag in both H and K.  
These data are presented in Table~\ref{tab:data} along with 
cross-identifications to previously published optical and infrared source lists.

In the optical (Optical ID column), Jones \& Walker (1988) numbers are listed 
with first priority, 
then Parenago (1954), Prosser et al. (1994), Hillenbrand et al (1997), 
and finally O'Dell \& Wong (1996) if no other designation exists. 
Several of the previously cataloged optical stars appear not to be real sources 
based on their absence in our NIRC images. Another, although unlikely,
possibility is that these are large-amplitude variables which have faded to
K$<$17.5 mag.  As listed in Hillenbrand (1997) these are 
459 and 699 (Jones \& Walker sources), 3071 and 3089 (Hillenbrand sources), and
9081 and 9326 (Prosser et al. sources).  One other object, 3083, is the head 
of a teardrop-shaped ``proplyd'' identified from high-resolution HST images; 
we have left this spatially extended source in our photometry list along with 
several other ``proplyds'' which may not be point sources (e.g. OW-114-426, also
extended in our images and clearly seen in silhouette in the Z-band data).
In the infrared (Alternate Infrared ID column) 
McCaughrean \& Stauffer numbers are given;
we recover all stars from that survey except for a few close pairs -- 
MS-86 is a close companion to P-1889, MS-65 is a close companion to P-1891 ($\Theta^1$C Ori), 
and MS-75 and MS-77 appear as a single source in our images.
Finally, $\sim$250 of the sources with K$<$14.5 mag in Table~\ref{tab:data} are also 
listed as point sources by 2MASS.

In summary, of the 778 stars in Table~\ref{tab:data}, $\sim350$ are previously 
known from optical studies conducted to varying survey depths while $\sim430$ 
are more heavily ``embedded'' in molecular cloud and/or circumstellar material.
Of the embedded sources, approximately $\sim125$ were previously catalogued 
(McCaughrean \& Stauffer over the inner 1.4' x 1.4'; Downes et al. 1981 and
Rieke, Low, and Kleinmann 1973 in the BN/KL region) and all those 
K $<$ 13.5 mag over the full area of the current NIRC survey were found in 
previous studies at lower spatial resolution and lower sensitivity 
(e.g. Hillenbrand et al. 1998). There are $\sim$175 sources newly catalogued 
here. Nearly all of these do appear in images available from McCaughrean 
or the NAOJ / Subaru Telescope first light press release.  

In deriving the ONC mass function, we have edited down the list of 778 in
Table~\ref{tab:data} to remove those with one or more of the following features:
1) no photometry at either K or H (21 sources);
2) photometry at K or H only, but not both (32 sources);
3) photometry given only as lower or upper limits at either K or H (38 sources);
4) photometry with internal errors $>$0.5 mag at either K or H (14 sources);
and
5) brighter counterparts in close pairs where we can not derive
photometry for the fainter counterpart because of contamination by
the brighter one (15 sources).  This last criteria was imposed so that 
the edited source listed would not be biased in any way against fainter, 
presumably lower mass, objects.  The number of stars remaining for our 
derivation of the ONC stellar/sub-stellar mass function is 658.

\section{Basic Results}

\subsection{The K Histogram and K--(K-H) Color-magnitude Diagram}

\begin{figure}
\vskip -0.5truein
\plotone{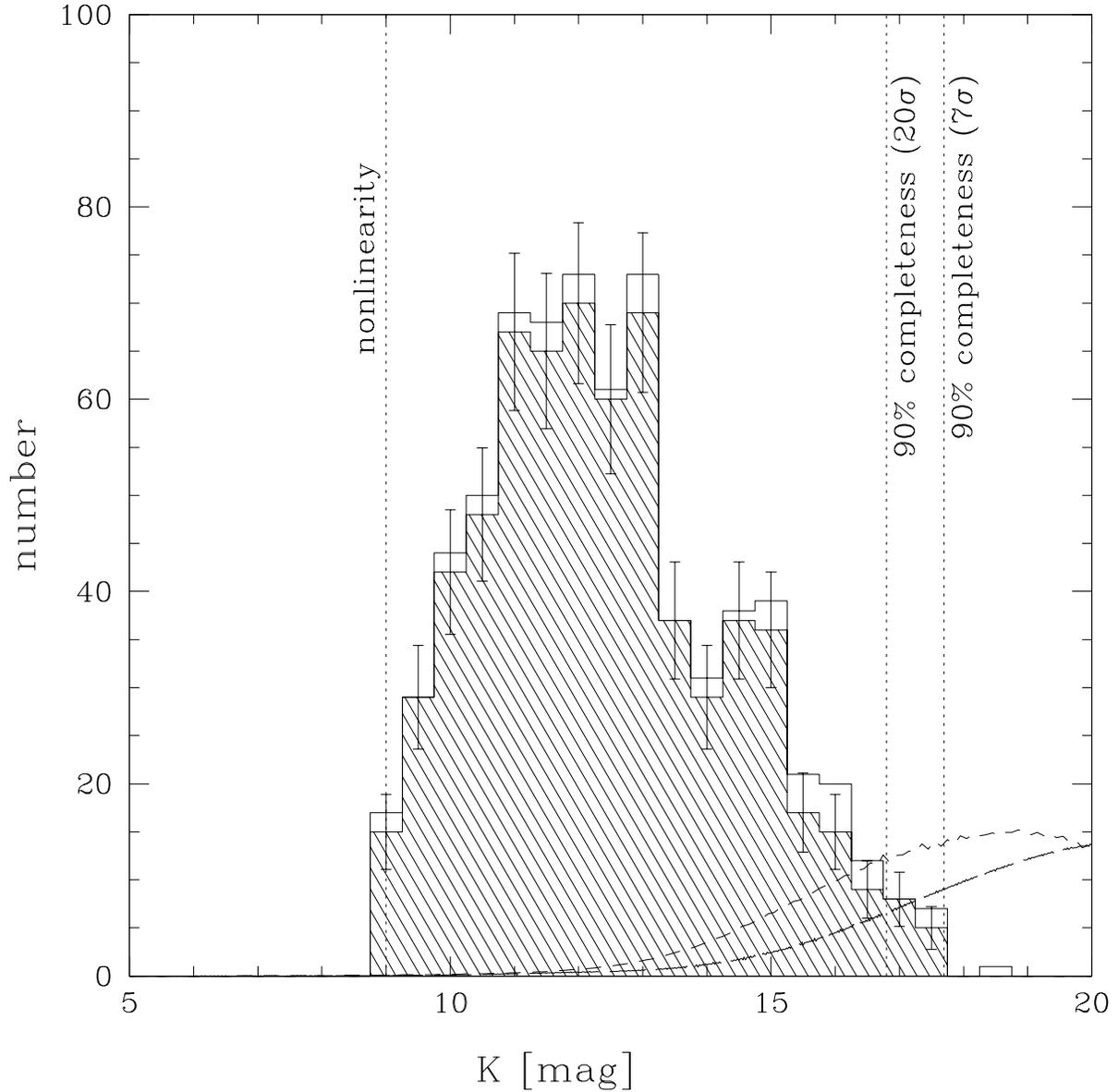}
\figcaption{Distribution of K magnitudes for stars photometered with NIRC.
The open histogram represent all stars with measured K magnitudes while
the hatched histogram represents a reduced sampled of stars used in the mass
function analysis.  See text for explanation of the second sample. 
Short-dashed line represents the Galactic model of
Wainscoat et al. (1992); long-dashed line represents the same model but
reddened for stars located behind the cloud by the extinction map shown
in Figure~\ref{fig:mosaics}.
\label{fig:khist}
}
\end{figure}

In Figure ~\ref{fig:khist} we present the histogram of K magnitudes for all 
objects with measurable photometry over our 5.1$'$ x 5.1$'$ field. 
Consistent with previous near-infrared studies of the ONC
(McCaughrean et al 1995; Ali \& DePoy 1995; Lada et al. 1996) we find
that the K magnitude histogram rises to a peak around K = 12-12.5 mag and
then declines.  The minor peak at K$\approx$14.5 mag
is also seen in Figure 2 of McCaughrean et al. (1995).
The hatched portion of the Figure shows the sample remaining after removing
stars without suitable photometry at both H and K according to the criteria
listed above.  This sample is not substantially different 
from the unhatched distribution representing the full photometric database.

\begin{figure}
\vskip -0.5truein
\plotone{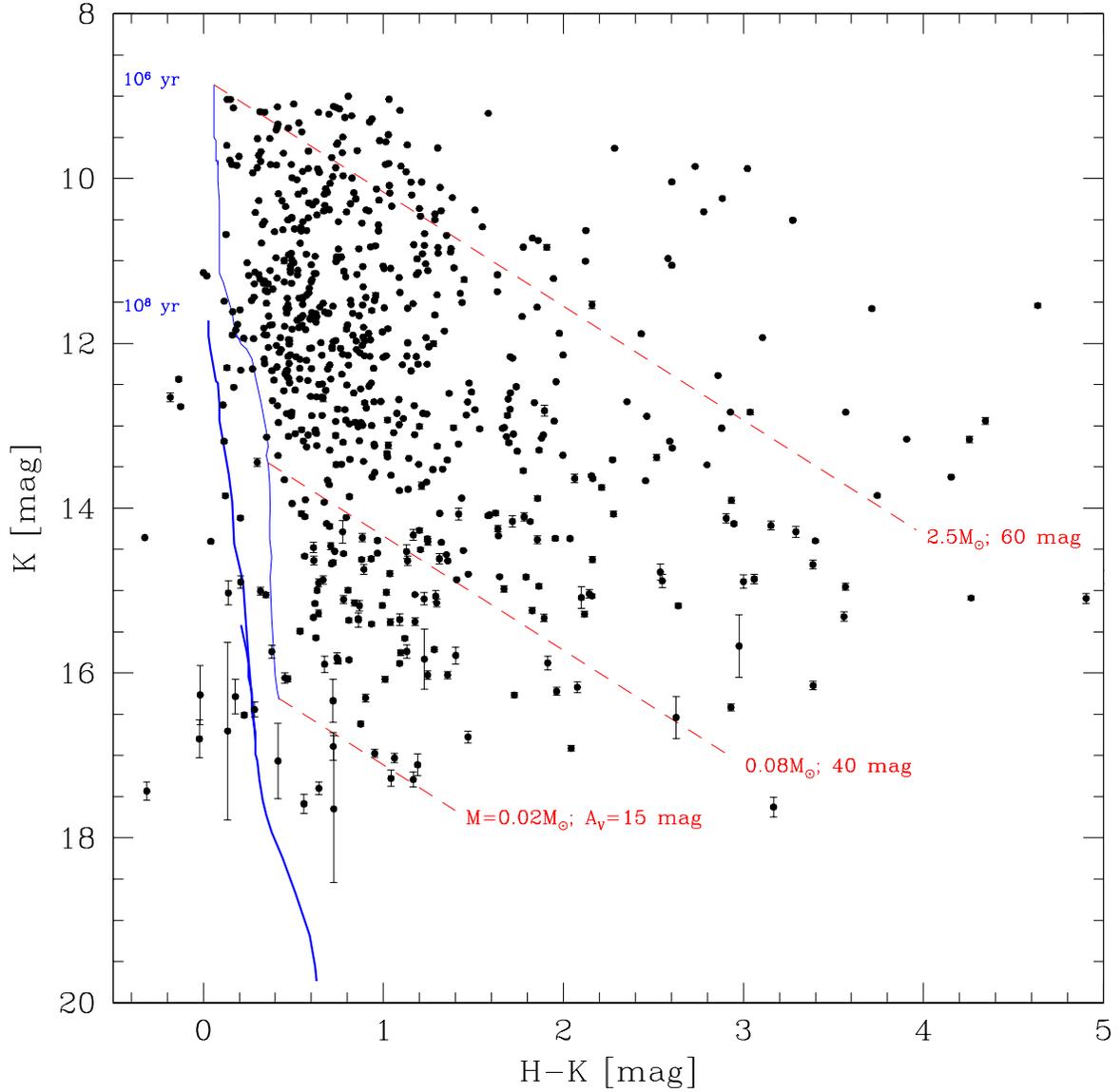}
\figcaption{K vs H-K diagram for stars photometered with NIRC.
Also shown is the 100 Myr isochrone (equivalent to the zero-age main
sequence for masses M$>$0.35 M$_\odot$) and the 1 Myr pre-main sequence
isochrone from D'Antona \& Mazzitelli (1997, 1998) translated into this
color-magnitude plane (solid lines). Reddening vectors (dashed lines) originate
from the 1 Myr isochrone at masses of 2.5 M$_\odot$, 0.08 M$_\odot$, and 0.02
M$_\odot$. We believe that the source detection is 90\% complete at the
7$\sigma$ theshold to K$>$17.5 mag. 
Internal errors in the K magnitudes are indicated; errors in the H-K 
color are larger than those in K band alone.  The limit for 10\% photometry 
occurs at K$\approx$17.3 mag and H$\approx$17.4 mag. 
\label{fig:khk.discrete}
}
\end{figure}

The \khk diagram for all stars with both H and K photometry from this study  
is presented in Figure~\ref{fig:khk.discrete}. The 100 Myr isochrone
(equivalent to the zero-age main sequence for masses M$>$0.35 M$_\odot$)
and the 1 Myr pre-main sequence isochrone from D'Antona \& Mazzitelli (1997, 1998),
translated into this color-magnitude plane as described in section 6.1.3, 
are shown.  Reddening vectors originating from the 1 Myr isochrone at masses of 
2.5 M$_\odot$, 0.08 M$_\odot$, and 0.02 M$_\odot$ are indicated.  
Considering only stellar photospheres for the moment,
our data are sensitive to all objects (stars and brown dwarfs)
with ages $\sim$1 Myr and masses M$>$0.02 M$_\odot$ seen through values 
of extinction A$_V <$ 10 mag.  The observed colors of most of the objects 
are substantially redder than the expectations from pre-main sequence 
isochrones, a fact which can be attributed to a combination of  
extinction and excess near-infrared emission due to a circumstellar disk
as discussed in section 6.2.  Nevertheless, several tens of
reddened objects located below the hydrogen burning limit at 0.08 M$_\odot$
are present.  Most of these are probable young brown dwarfs, although some
may be field stars.

\subsection{Field Star Contamination}


Our images and the resulting K-magnitude histogram and \khk 
color-magnitude diagram contain both ONC cluster members and unrelated 
field stars. Since H- and K-band photometry alone can not distinguish 
between cluster members and nonmembers,  we assessed contamination to the star
counts from field stars using a modified version the Galactic star count model 
of Wainscoat et al. (1992).  
While the nominal Wainscoat et al. model includes a smooth Galactic 
extinction distribution, the line of sight toward the ONC contains a substantial
and spatially variable extinction component from the Orion molecular cloud, as
was shown in Figure~\ref{fig:mosaics}c. 
This extinction map was generated from the
C$^{18}$O column density data of Goldsmith, Bergin, \& Lis (1997) by assuming
a C$^{18}$O/H$_2$ abundance of $1.7\times10^{-7}$ (Frerking, Langer, \& Wilson
1982) and that an H$_2$ column density of 10$^{21}$ cm$^{-2}$ corresponds to
1 magnitude of visual extinction (Bohlin, Savage, \& Drake 1978). The visual
extinction peaks along the western part of the inner ONC with a maximum value
A$_V$ = 75 mag, and falls off sharply to the east to a minimum value of 
A$_V$ = 3 mag at the edge of our NIRC map. 
Obviously the contribution from background field stars 
to the observed star counts will vary substantially across the ONC, 
in inverse relation to this extinction distribution. 
The number and near-infrared magnitudes and colors of expected field stars 
were obtained by convolving this extinction map with the nominal 
Wainscoat et al. star count model, although the additional extinction 
from the Orion molecular cloud was added only to the background 
field star population. 

The distribution of K magnitudes for the field stars was shown in 
Figure~\ref{fig:khist} (dashed curves), both before and after convolution 
with the molecular extinction map. With addition of the proper amount 
of extinction at the distance of the Orion cloud, the numbers predicted for
foreground/background contamination in our data are reduced to
0.08 stars arcmin$^{-2}$ at K$<$13 mag (from 0.17 stars arcmin$^{-2}$), 
0.27 stars arcmin$^{-2}$ at K$<$15 mag (from 0.76 stars arcmin$^{-2}$),
and 1.54 stars arcmin$^{-2}$ at K$<$18 mag (from 3.41 stars arcmin$^{-2}$). 
The total number of stars 
predicted to contaminate our NIRC photometry down to the K completeness limit 
(17.5 mag) is 34 (with 43 down to K=18 mag), representing a small but 
non-negligible 5\% of our survey sample.

In Figure~\ref{fig:khk.data_fieldstars} we compare the ONC data (panel a) to  
the model field star population (panel b).  For the data we have used the
edit source list of 658 stars discussed above and for the field stars we have
convolved the model with the photometric errors as a function of magnitude
characterizing the NIRC photometry. A Hess diagram format was adopted,
where individual points have been smoothed with an elliptical gaussian
corresponding to the photometric uncertainties. This Figure highlights
the large concentration of observed stars with K $\approx$ 12 and
H-K $\approx$ 0.5, and affirms that the the density of stars in the \khk
diagram is dominated by ONC cluster members at all but the faintest mangitudes.
To derive the \khk distribution of stars actually associated with the ONC 
we subtracted panel (b) from panel (a), as shown in 
Figure~\ref{fig:khk.tutorial}a. 

\section{Analysis: The ONC Mass Spectrum Across the Hydrogen-Burning Limit}

The goal of this study is to translate the information contained in the
\khk diagram shown in discrete format in Figure~\ref{fig:khk.discrete} 
and in Hess format in Figure~\ref{fig:khk.data_fieldstars}a,
into information on the stellar/sub-stellar mass function. This is not a trivial
transformation since the location of a young star in the \khk diagram depends
on four parameters: stellar mass, stellar age, presence and properties
(e.g. accretion rate) of a circumstellar disk, and extinction. A moderately
bright, red object, for example, while usually thought of as a massive star
seen through large extinction, can also be a much lower mass star with a large
near-infrared excess and significantly lower extinction. Indeed, bright red
objects can be reproduced by a number of combinations of stellar mass, 
stellar age, near-infrared excess, and foreground extinction.
Faint, blue stars on the other hand, can come only from the lower masses, 
older ages, smaller near-infrared excesses, and lower extinctions.  
The distribution of data points across the \khk diagram is dictated by 
the interaction occuring for each star in the cluster of these four 
primary physical parameters.

Photometry alone can not be used to deconvolve the age, near-infrared excess,
and extinction distributions to obtain uniquely the mass of the object. Such
an effort would require spectroscopic observations for the entire cluster
population, which currently do not exist. Therefore, a variety of techniques
making various assumptions about these parameters have been developed in
order to constrain the initial mass function. The most common of these 
approaches is to determine if the peak and the width of a K-band histogram 
are consistent with both an assumed initial mass function and a ``reasonable'' 
stellar age distribution (Zinnecker \& McCaughrean 1991). One drawback 
to this approach is that the information inherent to multi-band photometry 
often is not used to constrain other two parameters: the extinction and the
near-infrared excess properties of the individual stars.  
Other approaches attempt to use multi-band photometry to de-redden 
individual stars and to consider the effects of near-infrared excess in
estimating individual stellar masses from a single mean mass-luminosity 
relationship (e.g. Meyer 1996).
A drawback of this approach is that it assumes that the near-infrared excess 
at the shortest wavelength is negligible, and that the mean age of the entire 
cluster is a good approximation for each individual member of the cluster.

Here, we describe a new method for deriving the stellar mass function that
acknowledges the existence a distribution of stellar ages, near-infrared
excesses, and extinction values in star-forming regions.
We use these distributions to determine the probability 
a star could be a certain mass based on its location in the \khk diagram,
as opposed to estimating unique masses for individual stars.
The advantages to this approach are that we use all the photometric
information available, we make no a priori assumptions about the shape of the 
mass function, and we incorporate the inherent photometric uncertainties.
Before describing our method for inverting the observed \khk
diagram to derive the mass function, we first establish the stellar age 
and near-infrared excess distribitions appropriate for the ONC needed
for our analysis.

\subsection{Assumptions}

\subsubsection{Stellar Ages}

Hillenbrand (1997) used the D'Antona \& Mazzitelli (1994) tracks to find
a mean age for low-mass optically visible ONC stars of 0.8 Myr with 
an age spread of up to 2 Myr.  
We show in Figure~\ref{fig:agehist} the age distribution derived
using the more recent D'Antona \& Mazzitelli (1997, 1998) calculations and
updated spectral-type -- temperature -- color -- bolometric correction
relations, described below.  This Figure includes only stars within the area
of our NIRC survey.  Hillenbrand (1997) discussed the presence of a radial
gradient in the stellar ages where the mean age for stars in the inner ONC 
is slightly younger (by 0.25 dex or so) than the mean age for the ensemble ONC.
This trend is also present using the updated theory and 
observational-to-theoretical transformations.  
Based on Figure~\ref{fig:agehist}, we adopt in what 
follows an age distribution which is uniform in log between 10$^5$ and 10$^6$ yr;
we also consider a distribution which is uniform in log between 3x10$^4$ and 
3x10$^6$ yr with little difference in the results.

\subsubsection{Near-Infrared Excess}

We quantify the near-infrared excess using the H-K color excess, defined as    
$\Delta(H-K) = (H-K)_{observed} - (H-K)_{reddening} - (H-K)_{photosphere}$.
Spectroscopic and photometric data presented in
Hillenbrand (1997) and Hillenbrand et al. (1998) allow us to compute this 
quantify for those optically visible stars within the area of our NIRC mosaic. 
$(H-K)_{observed}$ is the tabulated color.
$(H-K)_{reddening} = 0.065~A_V$ where $A_V$ is derived from the spectral type 
and observed V-I color in comparison to expected V-I color and $A_V = 2.56~E(V-I)$. 
$(H-K)_{photosphere}$ comes from the relation between temperature and intrinsic
H-K color described below.
A histogram of the derived H-K excesses is shown in the top panel
of Figure~\ref{fig:irxshist}. We find that the observed near-infrared excesses
can be well represented by a half-gaussian with a dispersion $\sigma$=0.4 mag, 
as shown by the solid line.  In practice, we truncate the gaussian at 
$\Delta$(H-K) = 1 mag, which is the maximum H-K excess observed
in the inner ONC.
Hillenbrand et al. (1998) discussed the presence of a radial gradient in ONC 
near-infrared excess values with the mean near-infrared excess
(measured as $\Delta(I-K)$ instead of the $\Delta(H-K)$ used here)
larger for stars in the inner ONC than for the ensemble ONC.  We emphasize,
therefore, that the H-K excess distribution presented in 
Figure~\ref{fig:irxshist} is known to be accurate for the inner ONC only,
although we note that the distributions similarly calculated for young stellar
populations in Taurus-Auriga, IC348, L1641, NGC2264, NGC2024, MonR2, and
Chamaeleon using literature data (see description of samples and procedure
in Hillenbrand \& Meyer, 2000)
are generally similar in form although slightly narrower in width. 
  
In addition to the H-K excess, we must estimate the excess at K band alone 
in order to properly model the \khk diagram. Since the K magnitude excess is more
difficult to compute accurately than the H-K color excess, we have used the
K excesses tabulated for pre-main sequence stars in the Taurus molecular cloud 
by Strom et al. (1989) and the H-K excesses calculated as above using data from 
Kenyon \& Hartmann (1995) to establish
an empirical relation between these quantities. The bottom panel in
Figure~\ref{fig:irxshist} shows the correlation between the K and H-K excess
derived for stars in Taurus, which can be represented by a linear fit of
$\Delta$K = 1.785 $\times$ $\Delta$(H-K) + 0.134 with a scatter of $\pm$ 0.25
mag. We assume that this relationship also holds for stars in the ONC.

\subsubsection{Translations from Theoretical to Observational Quantities}

The final step before we can create models of the \khk diagram is conversion
of theoretical pre-main sequence evolution into the observational plane.
We use the theoretical description of luminosity and effective temperature 
evolution with mass according to D'Antona \& Mazzitelli (1997,1998).
These tracks are the only set available which cover the full range of masses 
sampled by our data, at the numerical resolution needed.  We note, however, 
that the most recently circulated calculations of pre-main sequence evolution 
by various groups (D'Antona \& Mazzitelli; Burrows et al.; Barraffe et al.;) 
do seem to be converging in the ranges where they overlap.  Nevertheless,
we must note that the details of our results likely are sensitive 
to the set of tracks/isochrones we have adopted.

We have transformed the D'Antona \& Mazzitelli (1997,1998) calculations of 
L/L$_\odot$ and T$_{eff}$/K into K magnitude and H-K color using Chebyshev fits 
(Press et al. 1989) to bolometric correction,
V-I color, I-K color, and H-K color vs effective temperature.  
For the mass range of interest in this paper we have taken the empirical data 
on bolometric corrections from Bessell (1991), Bessell \& Brett (1988), 
and Tinney, Mould, \& Reid (1993);
on colors from Bessell \& Brett (1988), Bessell (1991), Bessell (1995),
Kirkpatrick \& McCarthy (1994), and Leggett, Allard, and Hauschildt(1998); and
on effective temperatures from Cohen \& Kuhi (1979) 
--  effectively Bessell (1991) --
Wilking, Greene, \& Meyer (1999), and Reid et al. (1999).

Note that these relationships are somewhat different than those used in
Hillenbrand (1997, 1998). We have now shifted the temperature scale cooler
and the bolometric corrections slightly smaller at the latest spectral types,
in keeping with current consensus that was not well-established at the time
of our earlier work.  The combination of updated tracks and 
updated transformations between observations and theory 
have caused a shift in our interpretation of the optical data presented 
by Hillenbrand (1997).   As we show in Figure~\ref{fig:optical.newvsold}, 
instead of a mass function that rises to a peak, flattens, and shows evidence 
for a turnover (bottom panel), the same data now appear to suggest a mass 
function for the greater ONC which continue to rise to the mass limit 
of our previous survey (top panel).

Finally, in the current analysis,
we use the Cohen et al. (1981) reddening vector, A$_K$=0.090~A$_V$ 
and A$_H$=0.155~A$_V$, 
and assume the Genzel et al. (1981) distance of 480$\pm$80 pc to the ONC 
(distance modulus = 8.41 mag). 

\subsection{A Model for the Distribution of Stars in the \khk Diagram}


Given the above assumptions concerning the age and near-infrared excess
distributions for the inner ONC, and the translation of theoretical
tracks/isochrones into the \khk plane, we are now in a position to construct
model \khk diagrams. In Figure~\ref{fig:khk.tutorial} we illustrate the
effects of various age and near-infrared excess distributions on the appearance
of the \khk diagram.  Panel (a) shows discrete isochrones from the
calculations of D'Antona \& Mazzitelli (1997, 1998) for ages of 10$^5$ and
10$^6$ year; panel (b) shows a sample of stars uniformly distributed in log-mass
between 0.02-3.0 M$_\odot$ and uniformly distributed in log age between
10$^5$ and 10$^6$ year; panel (c) shows the same mass and age distribution
of (b) but now includes the near-infrared excess distribution
parameterized in Figure~\ref{fig:irxshist}.  No extinction is included
in these panels.  Note that in the case of a uniform age distribution
(panel b), the \khk diagram is {\it not} uniformly populated between the
limiting isochrones. As originally shown by Zinnecker \& McCaughrean (1991) in
an analysis of K band histograms, the onset of deuterium burning occurs at
different times for different masses, leading to distinctive peaks in the
magnitude and color-magnitude distribution for pre-main sequence stars. These
peaks become less distinctive when a near-infrared distribution is added (as
shown in panel c) and even less distinctive when extinction is added (as shown
next).

\begin{figure}
\vskip -0.5truein
\plotfiddle{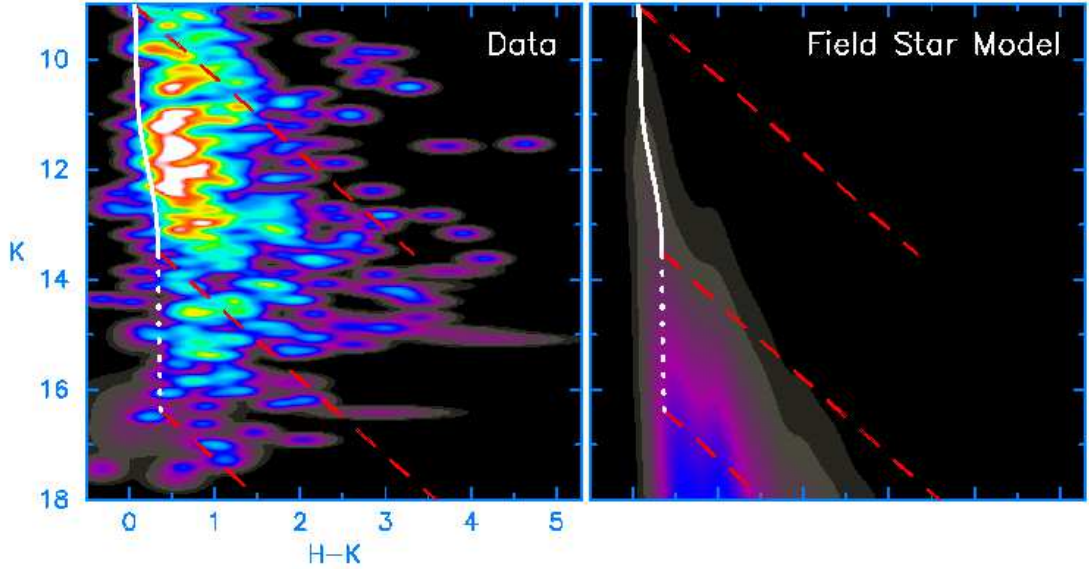}{5.in}{-90}{70}{70}{-260}{400}
\figcaption{Hess format \khk diagram for our data (left panel) and 
an appropriately reddened field star model (right panel).
To generate the contours for the observations, individual stars 
were smoothed by an elliptical gaussian corresponding to their 
photometric errors as described in the text. Similarly, the field star model
was convolved with the typical photometric error as a function of magnitude. 
The white solid/dotted line is the 1 Myr pre-main sequence isochrone 
with the transition from a solid to dotted occuring at the hydrogen burning 
limit of 0.08 M$_odot$. The lowest mass represented by the isochrone 
is 0.017~M$_\odot$. The reddening vector for 
A$_V <$ 50 mag is indicated by red dashed lines. The color stretch is
identical for both panels, with the data plot containing 658 stars and the
field star model containing 34 stars down K=17.5
and 43 stars down to K $<$ 18 mag. These figures demonstrate that
field stars make a negligible contribution to the ONC star counts except
at K $>$ 16 mag (see also Figure~\ref{fig:khist}); by K $>$ 17 mag the
field stars dominate cluster members.
\label{fig:khk.data_fieldstars}
}
\end{figure}

\begin{figure}
\vskip -1.5truein
\plotone{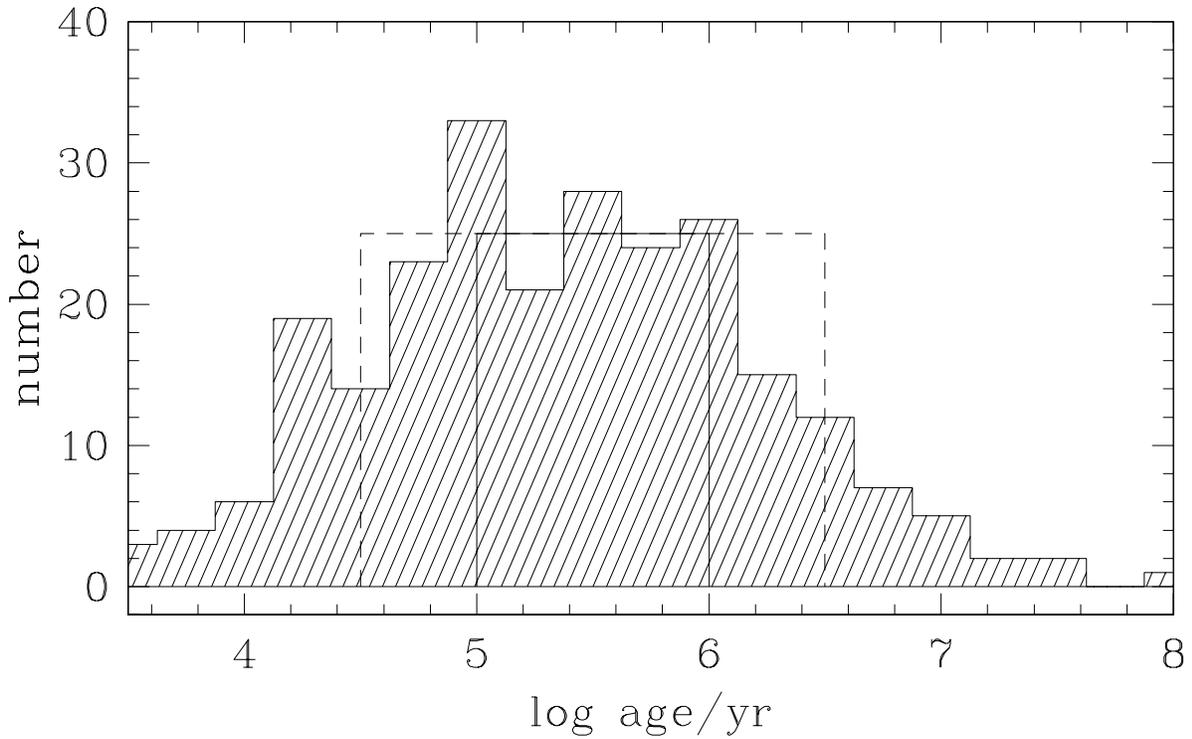}
\vskip -0.25truein
\figcaption{Distribution of ages for optically visible ONC stars with 
M $<$1.5 M$_\odot$ located within the boundaries of our NIRC mosaics. 
This Figure was constructed using the data in Hillenbrand (1997) but the 
transformations between observational and theoretical quantities, and the
pre-main sequence evolutionary calcuations adopted in this paper.  
For the current analysis we assume an age distribution which is uniform 
in log between 10$^5$ and 10$^6$ yr, shown as the solid line, and we also 
consider an age distribution which is uniform in log between 
3x10$^4$ and 3x10$^6$ yr, shown as the dashed line.
\label{fig:agehist}
}
\end{figure}

\begin{figure}
\epsscale{0.85}
\vskip -0.6truein
\plotone{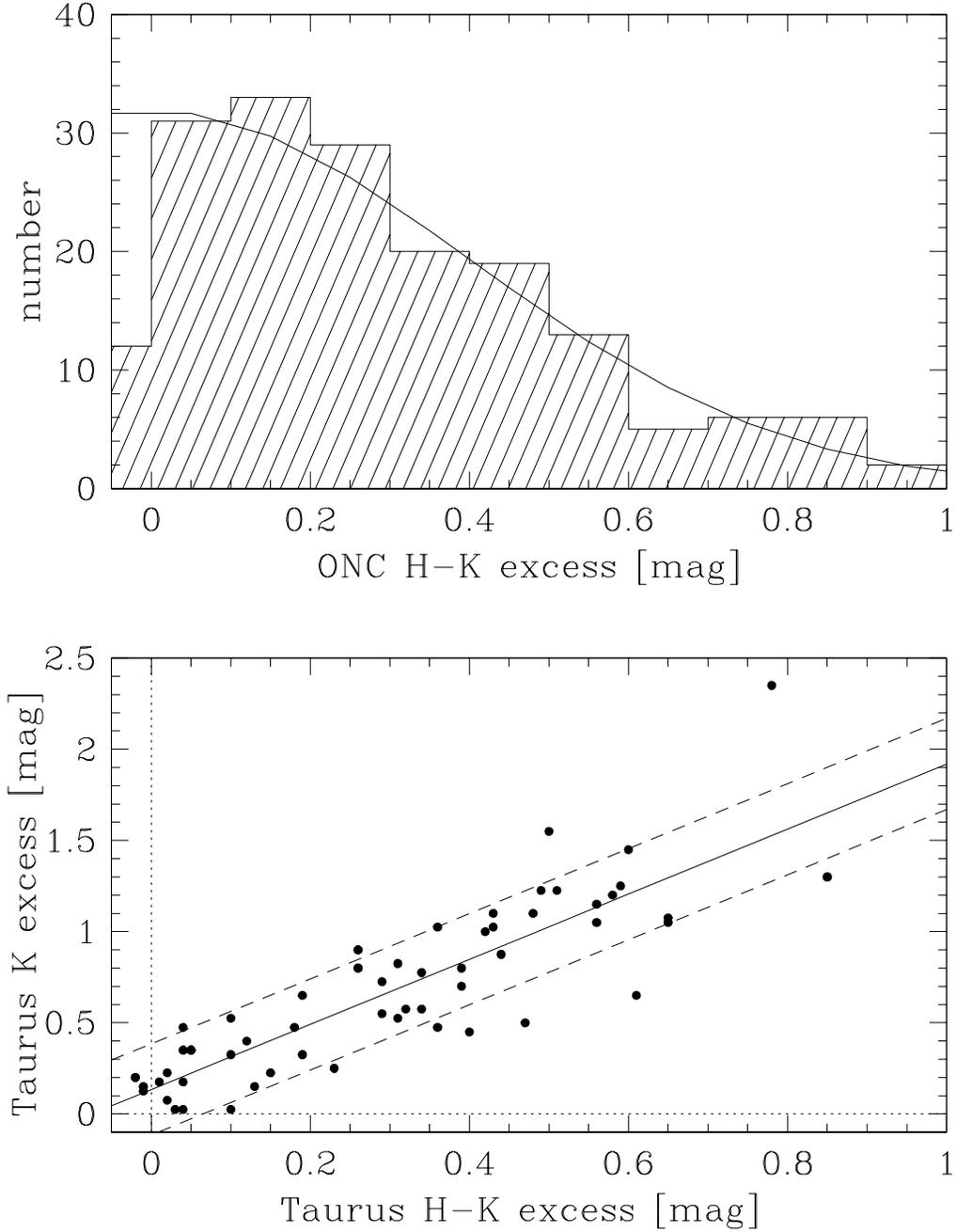}
\vskip -0.25truein
\figcaption{Distribution of K and H-K excesses. The top panel shows a
histogram of H-K color excesses for ONC stars located within the field of view 
of our NIRC mosaics, calculated using data from Hillenbrand et al. (1997, 1998).
The solid curve is a half-gaussian fit to the distribution and has
a dispersion $\sigma$=0.4 mag. The bottom panel shows the
correlation between K band excess and H-K color excess for stars in Taurus,
calculated using data from Strom et al. (1989) and Kenyon \& Hartmann (1995).
The solid line is the best fit to these data, $\Delta$K = 1.785 $\times$
$\Delta$(H-K) + 0.134 with the dashed lines indicating $\pm$ 0.25 mag scatter. 
In analyzing the ONC mass function we assume the distribution of H-K excess 
shown in the top panel, and the K band excess correlation with H-K excess 
shown in the bottom panel.
\label{fig:irxshist}
}
\epsscale{0.85}
\end{figure}

\begin{figure}
\epsscale{0.85}
\vskip -0.5truein
\plotone{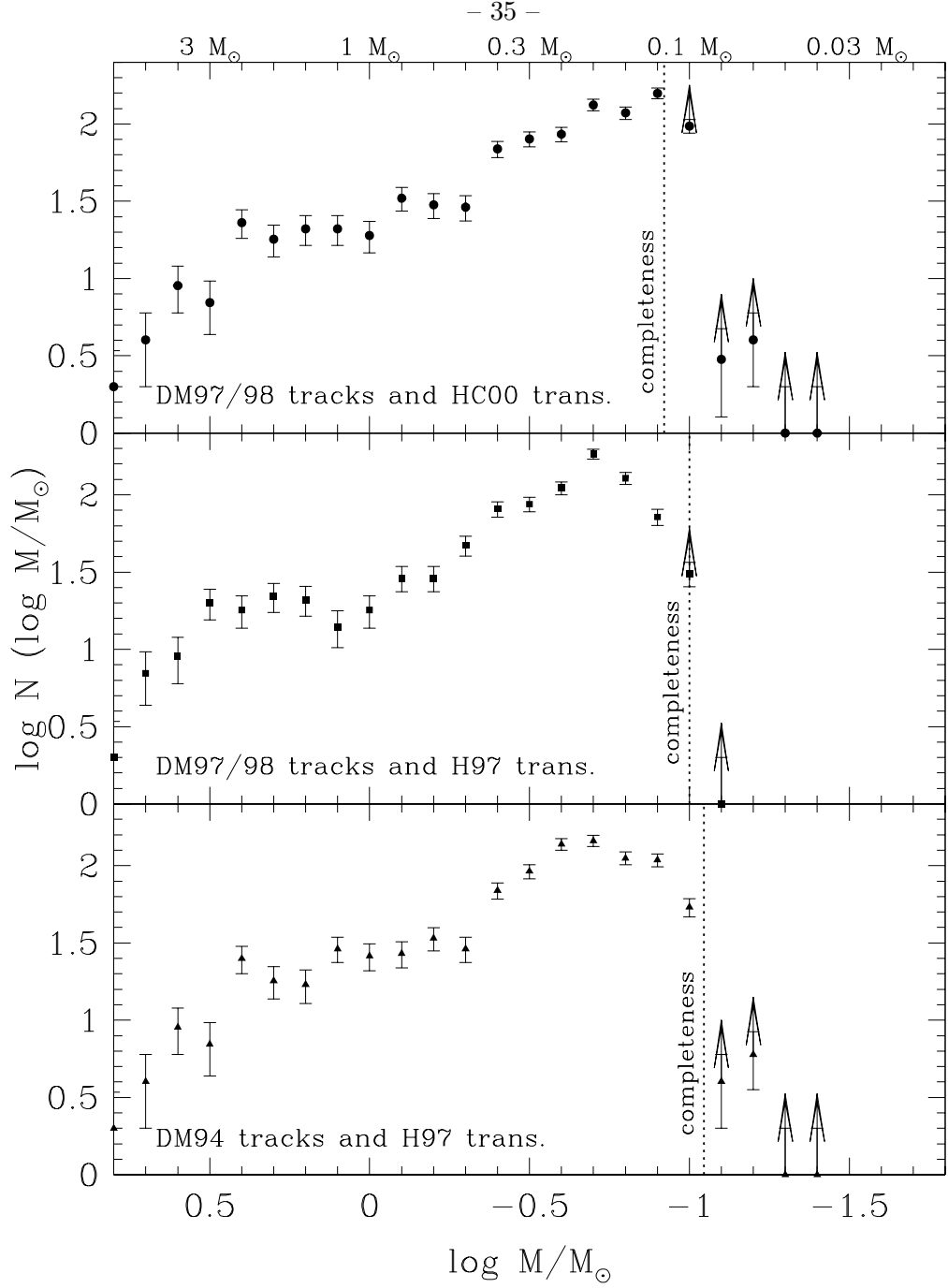}
\figcaption{ONC mass spectrum derived using the {\it optical} data 
of Hillenbrand (1997).  The input photometry and spectroscopy are the same 
in all three panels, and represent stars over 30$'$ x 34$'$ of the ONC.
In the top panel we show the mass function produced by the theoretical
description of luminosity and effective temperature evolution with mass
of D'Antona \& Mazzitelli (1997,1998) and the transformations between
observational and theoretical quantities adopted in this paper. 
In the middle panel we show the same tracks with the observational-theoretical
calibrations adopted by Hillenbrand (1997).
In the bottom panel we show the mass function produced by the
D'Antona \& Mazzitelli (1994) calculations and the 
calibrations adopted by Hillenbrand (1997).  
Note the dramatic difference in shape of the mass function 
below 0.2 M$_\odot$ between these three panels.
\label{fig:optical.newvsold}
}
\epsscale{1.00}
\end{figure}

\begin{figure}
\vskip -0.5truein
\plotfiddle{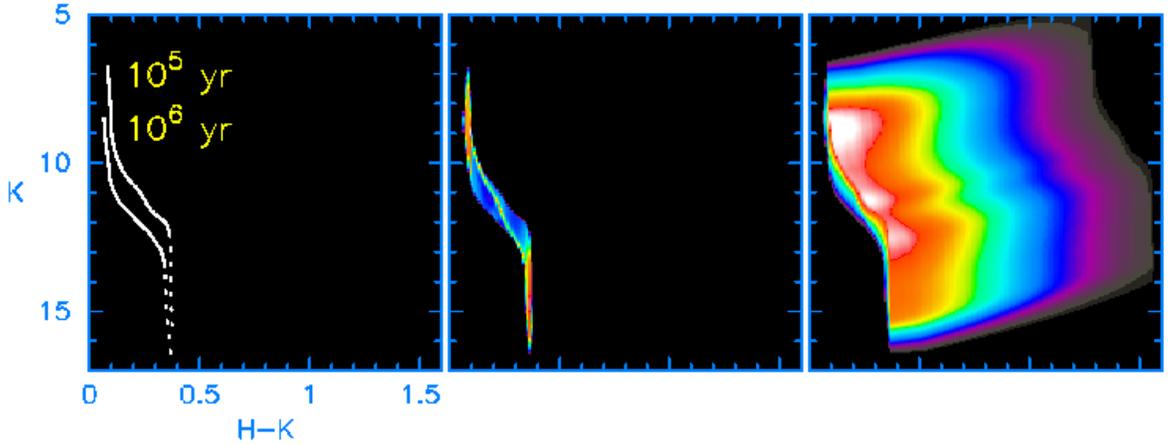}{5.in}{-90}{75}{75}{-260}{400}
\figcaption{
Model \khk diagrams for various assumptions about the age and near-infrared
excess distributions.  The mass function is log-uniform between
0.017 and 3.0 M$_\odot$. The left panel shows the \khk distribution of two 
single-aged populations at 10$^5$ and 10$^6$ years, 
with no near-infrared excess. The middle panel shows a population 
distributed log-uniform in age between 10$^5$ and 10$^6$ years, as we adopt
for the ONC (see Figure~\ref{fig:agehist}), and again with no near-infrared excess.
The right panel shows the same log-uniform age distribution but now includes
the near-infrared excess distribution adopted for the ONC 
(see Figure~\ref{fig:irxshist}).
\label{fig:khk.tutorial}
}
\end{figure}

\begin{figure}
\vskip -0.5truein
\plotfiddle{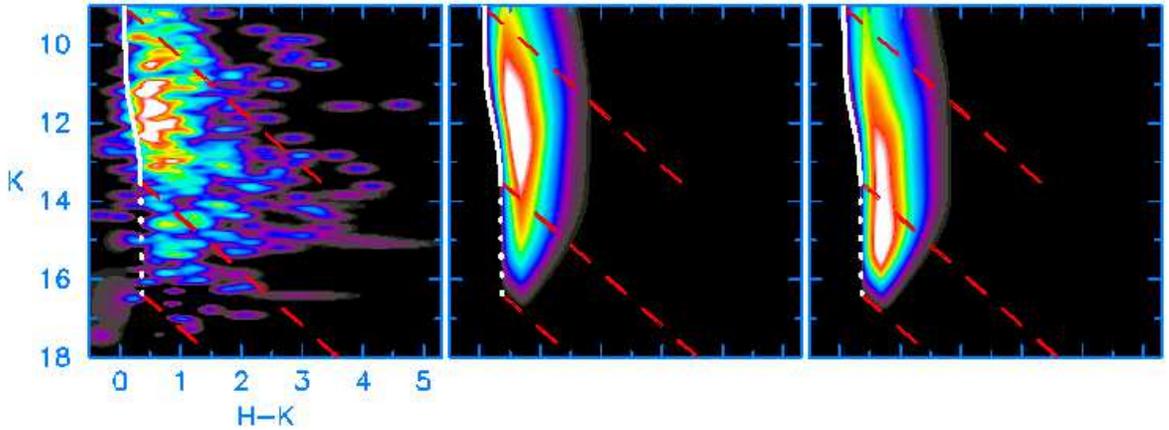}{5.in}{-90}{75}{75}{-260}{400}
\figcaption{Simulations of the \khk diagram using the age distribution
assumed from Figure~\ref{fig:agehist}, the near-infrared excess distribution assumed
from Figure~\ref{fig:irxshist}, and an extinction distribution which is uniform
in the interval A$_V$=0-5 mag.  The middle
panel shows the log-normal form of the Miller-Scalo mass function
while the right panel shows a shallow power law mass function 
(N(log M) $\propto$ M$^{-0.35}$).  Our data are shown in the left panel,
which is the subtraction of the field star model in 
Figure~\ref{fig:khk.data_fieldstars}b from the observations in 
Figure~\ref{fig:khk.data_fieldstars}a.  The models
suggest that a falling mass function like that of Miller-Scalo better 
represents the peak in the observed ONC star counts than does
an increasing mass function like the shallow power-law.  Although there 
appear to be some more highly extincted stars in the data than in
these models, broadening the A$_V$ distribution in the models dilutes the 
peak; this suggests that the bulk of the ONC stars are found at relatively low 
extinction, A$_V <$ 10 mag.
\label{fig:khk.ms_vs_pow}
}
\end{figure}

We incorporate elements of Figure~\ref{fig:khk.tutorial} to show in 
Figure~\ref{fig:khk.ms_vs_pow} (note the change in scale, now set to match
the range of our data) two model \khk diagrams in comparison to our NIRC 
observations. Panel (a) shows the ONC data of Figure~\ref{fig:khk.data_fieldstars}a
with the field star model of Figure~\ref{fig:khk.data_fieldstars}b subtracted. 
Panel (b) shows a stellar population distributed in mass according to the
Miller-Scalo mass function and distributed in age between 10$^5$-10$^6$ yr 
log-uniform, then also having the near-infrared excess distribution 
parameterized in Figure~\ref{fig:irxshist} and seen through extinction 
uniformly distributed between A$_V$=0-5 mag.
Panel (c) is the same as panel (b) except that the mass distribution is now
a power-law function instead of a Miller-Scalo function. 
The salient difference between these two mass functions is that 
the Miller-Scalo function (N(log M) $\propto$ e$^{-C1(log M - C2)^2}$; 
C=1.14, C2=-0.88 as in Miller \& Scalo, 1979) slowly declines across the hydrogen burning 
limit as N(log M) $\propto$ M$^{0.37}$ if forced to a power-law,  
while the straight power-law function 
(N(log M) $\propto$ M$^{-0.35}$) slowly rises. 
In creating Figure~\ref{fig:khk.ms_vs_pow} we have not attempted 
to reproduce the observations; we wish merely to illustrate the combined 
effects in the \khk diagram of different assumptions about the mass, age, 
near-infrared excess, 
and extinction distributions. Note in particular that there are many stars 
observed (panel a) through higher values of A$_V$ than we have 
considered in the models (panels b and c). Nevertheless, if we accept 
that our assumptions about the age and near-infrared excess distributions  
(as derived from optically visible stars in exactly this region) 
are approximately correct, then we must conclude that a declining 
mass distribution such as the Miller-Scalo function is a much 
better match to the data than a rising power-law (or even a flat) function. 
We quantify these impressions in the following section.  

\subsection{Implementation and Tests}

\subsubsection{Calculating Mass Probability Distributions} 

How can the effects of stellar age, near-infrared excess, and
extinction be disentangled to derive the mass function?  As already discussed,
the main difficulty is that more than one stellar mass can contribute power 
to any particular location in the \khk diagram through various combinations
of these variables. Fortunately, however,
the range of stellar masses that a given H-K,K data point could represent
is constrained by the stellar age and near-infrared excess distributions, 
which for the inner ONC we are able to measure (Figures~\ref{fig:agehist} and 
~\ref{fig:irxshist}), and by the slope of the reddening vector.  

In practice we calculate the stellar mass function as follows.
We take as a starting point the observed \khk grid shown in 
Figure~\ref{fig:khk.ms_vs_pow}a.  Recall that each star has been 
smeared out in this diagram by an elliptical gaussian corresponding 
to its photometric error; increasing the error even by a factor of 3
in each direction does not change the form of the distribution.  
We project every 0.01 mag wide pixel populated 
by data back along the reddening vector to establish which 
of the other pixels are crossed, and hence which combinations of unreddened
K magnitudes and H-K colors the star or star-plus-disk system could have.
Using a model \khk diagram, we keep track of the probability that a star 
of given mass can occupy that H-K,K combination given the
assumed stellar age and near-infrared excess distributions. 
We sum the probabilities for all of the possible H-K,K combinations
along the reddening vector, and then normalize to unity the integerated 
probability over all masses 0.02-3.0 M$_\odot$; i.e., the star must have 
have some mass within the considered range. 
By weighting the mass distribution derived for each
pixel in Figure~\ref{fig:khk.ms_vs_pow}a by the relative density of
observational data it represents, and summing the probability distribution
for all pixels, we produce the cluster mass function. In a similar manner,
we calculate probability distributions in A$_V$ for each pixel, which we
also density weight and sum to produce the cluster extinction distribution.

\begin{figure}
\epsscale{0.85}
\plotfiddle{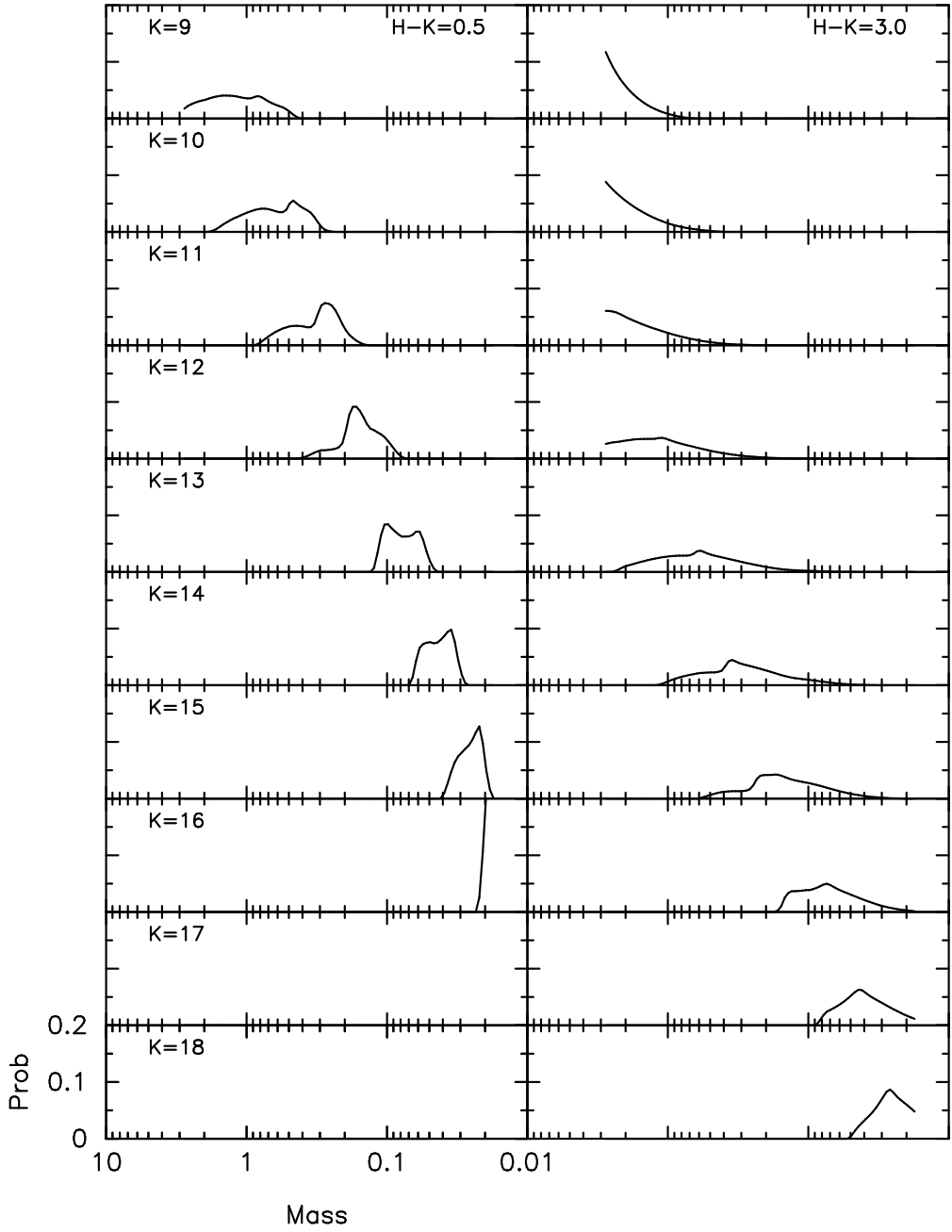}{7.in}{0}{75}{75}{-230}{000}
\figcaption{Illustrative mass probability functions derived
using our methodology.  Left panels show stars with H-K=0.5 and
right panels show stars with H-K=3.0, both columns of panels decreasing 
in brightness top to bottom from K=9 to K=18. The de-reddening model
uses the same distributions in age and in near-infrared excess
as employed elsewhere in this paper.  Note the tails upward at the 
lower and upper mass extrema in the panels for K=16, H-K=0.5 and K=9, 
H-K=3.0, respectively. These are caused by our imposition
of integrated probability equal to unity over the mass range 0.02-3.0 M$_\odot$. 
\label{fig:massprob}
}
\epsscale{1.00}
\end{figure}

Examples of individual stellar mass probability distributions obtained by 
de-reddening a star in the \khk diagram are shown in Figure~\ref{fig:massprob}
for a representative set of K magnitudes and H-K colors. 
A K=15 mag relatively blue (H-K=0.5 mag) star 
is permitted to have a mass anywhere in the range $\sim$0.02-0.04 M$_\odot$ 
with a most likely value just above 0.02 M$_\odot$, while a K=15 mag much redder
(H-K=3.0 mag) star has a broader range of permitted masses, 
$\sim$0.03-0.6 M$_\odot$ with a most likely value $\sim$0.2 M$_\odot$.
Note the tails upward at the lower and upper mass extrema in the panels for 
K=16, H-K=0.5 and K=9, H-K=3.0, respectively. These are caused by our 
imposition of integrated probability equal to unity over the mass range 
contained in the theoretical grid; in reality such stars have some probability 
of coming from smaller and higher masses (respectively) 
than the 0.02-3.0 M$_\odot$ range considered here.  The requirement of
an integrated mass probability of unity means that the outer few bins
of our resultant mass distribution may be unreliable.

Examples of individual extinction probability distributions are not shown
since extinction is essentially the independent variable in our technique.
Because we project each ``star'' back along the reddening vector and keep
track of the stellar mass, stellar age, and near-infrared excess combinations
that can conspire to produce that color-magnitude location, any given star 
can have any value of extinction ranging from a minimum of zero (in general,
although it is not always true that there is a zero-extinction solution) 
to a maximum set by the case of de-reddening to the oldest considered age 
(i.e. bluest possible original location in the \khk diagram) 
and having no near-infrared excess.  The result is that
an extinction distribution which is uniform will be recovered 
using our methodology as an extinction distribution that has an extended tail
induced by a combination of the age range and the near-infrared excess range 
considered in the de-reddening process.

\subsubsection{Tests of Methodology} 

\begin{figure}
\epsscale{0.85}
\vskip -0.5truein
\plotone{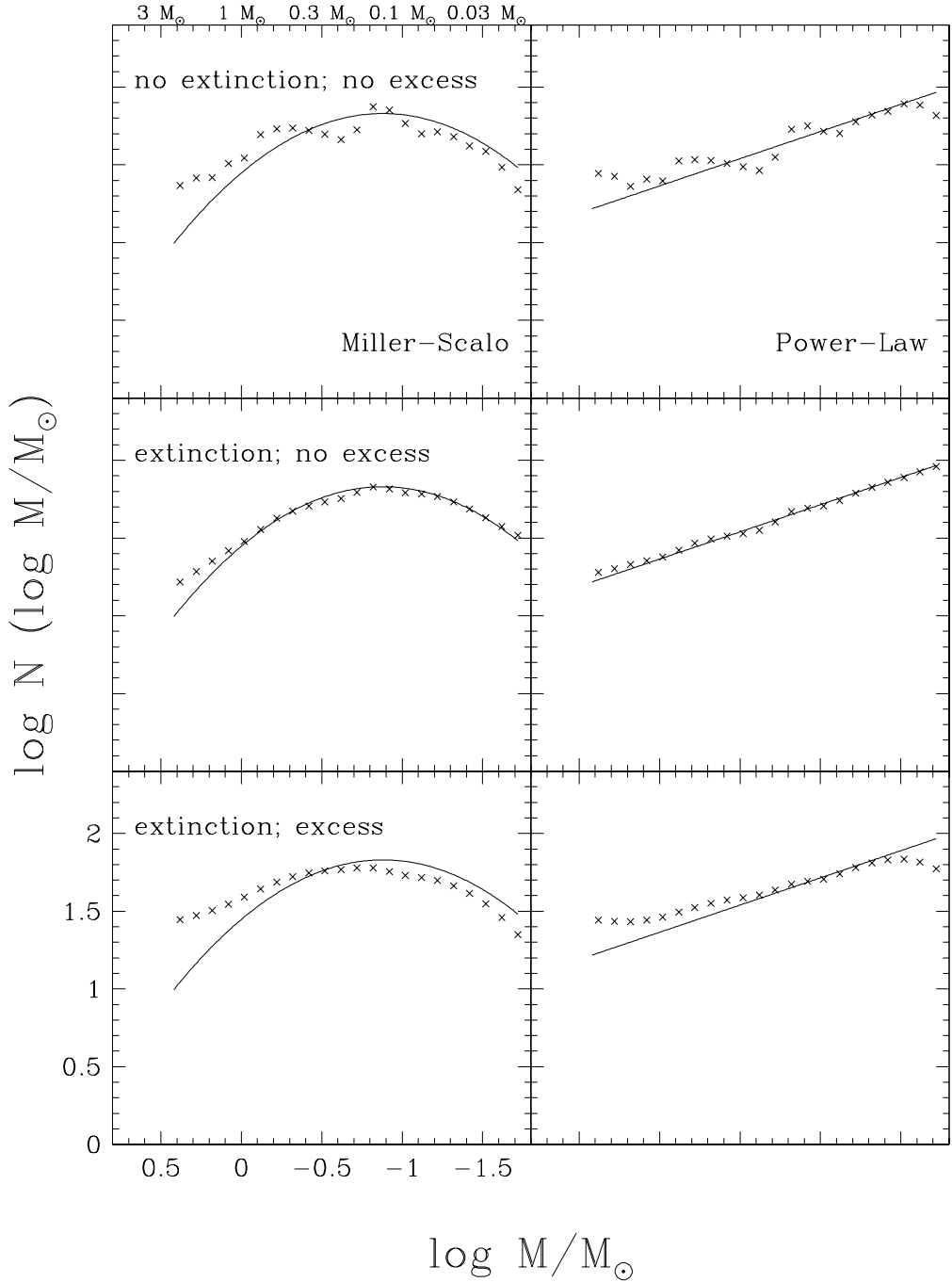}
\figcaption{Tests of the ability of our method to recover an input mass 
function.  Solid lines represent the input mass function while crosses represent
the recovered mass function.  Tests using the Miller-Scalo mass function appear
in the left panels and those using a shallow power-law mass function 
N(log M/M$_\odot$) $\propto$ (M/M$_\odot$)$^{-0.35}$ 
in the right panels; the age distribution in both the left and right panels
is log-uniform between 10$^5$ and 10$^6$ years. From top to bottom the panels 
indicate a) no extinction and no near-infrared excess;
b) extinction uniformly distributed A$_V$=0-5 mag and no near-infrared excess; 
and c) extinction uniformly distributed between A$_V$=0-5 mag and
near-infrared excess distributed using the half-Gaussian function described
elsewhere.  In every case we are able to distinguish between the 
slowly falling and the slowly rising mass functions.  
\label{fig:testmodels}
}
\epsscale{1.00}
\end{figure}

\begin{figure}
\epsscale{0.85}
\vskip -0.5truein
\plotone{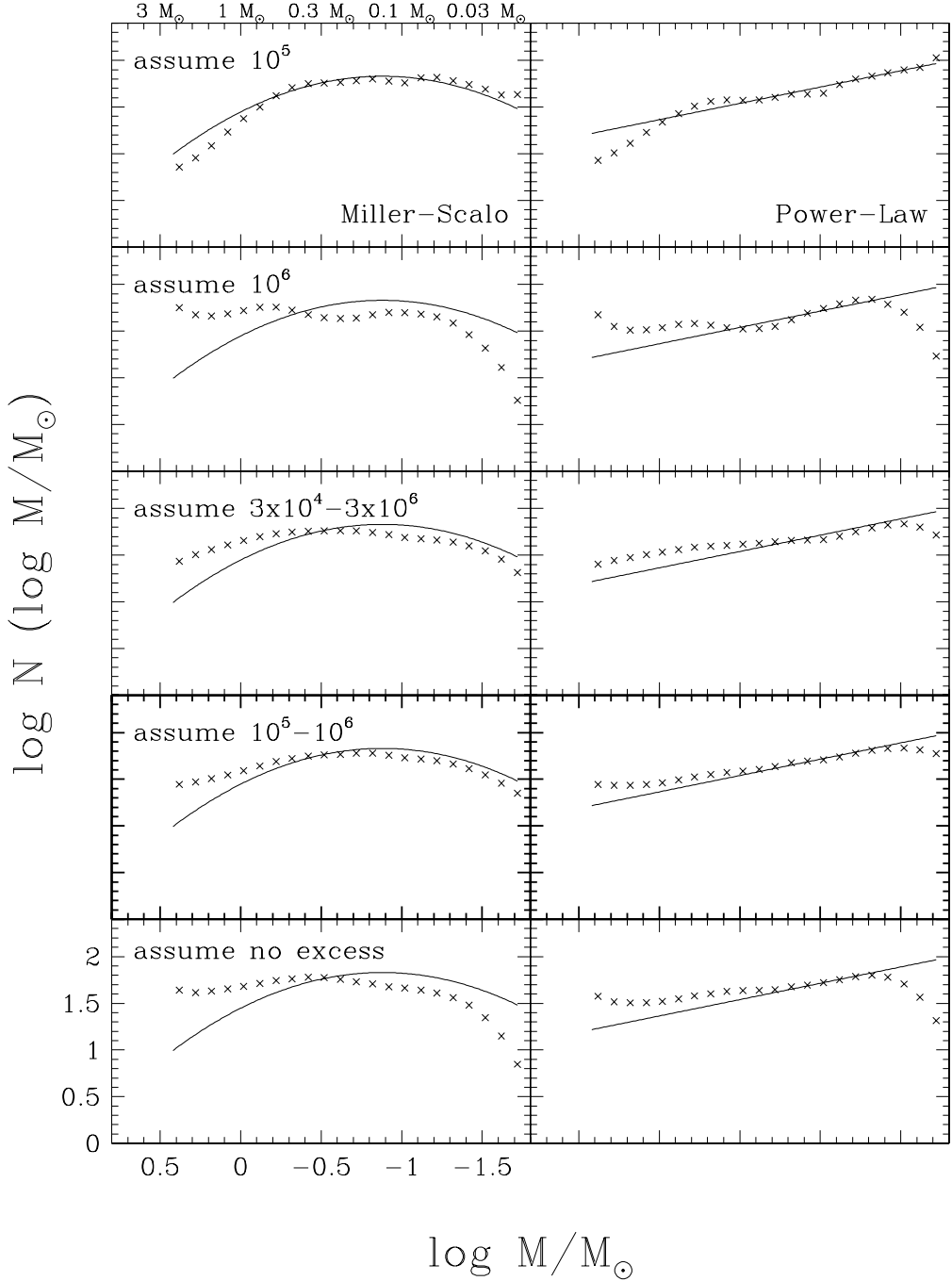}
\vskip -0.1truein
\figcaption{Tests of the ability of our method to recover an input mass 
when we intentionally assume an incorrect age or near-infrared distribution.
Solid lines represent the input mass function while crosses represent
the recovered mass function.  The Miller-Scalo mass function is tested
in the left panels while a shallow power-law mass function 
N(log M/M$_\odot$) $\propto$ (M/M$_\odot$)$^{-0.35}$ 
is tested in the right panels. In all panels the input age distribution 
is log-uniform between 10$^5$ and 10$^6$ years, the input near-infrared 
excess distribution is the half-Gaussian function discussed elsewhere,
and the input extinction distribution is uniform
between A$_V$=0-5 mag.  From top to bottom we have varied the assumptions in
recovering the mass functions to test incorrect ages (10$^5$, 10$^6$, and
log-uniform between 3x10$^4$ and 3x10$^6$ years), and to
test an incorrect near-infrared excess assumption (no infrared excess).
For reference, we also show in the fourth set of panels from top,
the results when the correct age and the correct near-infrared excess 
distributions are assumed.
\label{fig:testassumptions}
}
\epsscale{1.00}
\end{figure}

Before applying our newly developed methodology for deriving the 
stellar/sub-stellar mass function, we wish to test how accurately this method
can recover a known mass function.  For these tests we generated cluster
models with various stellar mass, stellar age, near-infrared excess, 
and extinction distributions, and then attempted to recover the underlying
mass function using the procedure described above. Figures~\ref{fig:testmodels} 
and ~\ref{fig:testassumptions} illustrate a sampling of the results.
In general, we are fairly confident in our ability to recover the general form 
of the input mass distribution for masses 0.03 $<$ M/M$_\odot$ $<$ 1.
Outside of these mass limits we suffer problems due to ``edge effects'' 
given the 0.02-3.0 M$_\odot$ range of the theoretical models 
we employ, and also due to saturation in our data at K $<$ 9 mag.  
In particular we note that in all cases we easily distinguish between 
mass functions that slowly fall across the hydrogen burning limit into the
brown dwarf regime, as N(log M) $\propto$ M$^{0.37}$, 
and those that slowly rise, as N(log M) $\propto$ M$^{-0.35}$.

In Figure~\ref{fig:testmodels}, we present test results where the stellar age
and near-infrared excess distribution assumed in extracting the mass function from the
\khk diagram is the same as the input cluster model. Thus these tests probe
the success of the method when the cluster properties are accurately known 
a priori.  Each of
these models contain a log-uniform age distribution between 10$^5$-10$^6$ yr.
The left panels are for models where the input mass function is Miller-Scalo
while the right panels are for a power law of form 
N(log M) $\propto$ M$^{-0.35}$. The top panels are models with no extinction
and no near-infrared excess; the middle panels include uniform extinction
between A$_V$=0-5 mag; and the bottom panels include both
uniform extinction and the near-infrared excess distribution parameterized in
Figure~\ref{fig:irxshist}. Note that the two bottom panels correspond to the 
cases shown in the \khk diagrams of Figure~\ref{fig:khk.ms_vs_pow}bc.
Looking at the difference between the top and
middle panels, addition of extinction to a model (surprisingly) helps our
method to recover the input mass function. Looking at the difference between
the middle and bottom panels, addition of near-infrared excess hurts slightly
but only at the tails in mass. These test results illustrate that our method
can never {\it perfectly} recover the input mass function as long as there is a
spread of ages or of near-infrared excesses --
even if these distributions are known.  H- and K-band photometry alone
can not uniquely determine the mass, age, near-infrared excess, and extinction
which go into producing the observed color-magnitude location, 
and hence our method considers all possible combinations of these parameters. 
The result is imperfect; nevertheless, it seems clear from
Figure~\ref{fig:testmodels} that we do reasonably well in recovering the general
shape of the input mass function.

In Figure~\ref{fig:testassumptions}, we present test results where we
deliberately choose an incorrect cluster age and/or near-infrared distribution
to de-redden the \khk diagram. These tests probe how robust our procedure is in
recovering the mass function when faced with uncertainties in characterizing 
the actual cluster properties. In each of these tests, the input cluster 
contains the same age distribution and near-infrared excess distribution 
that we assumed for the ONC. In addition, we added uniform extinction between 
A$_V$=0-5 mag. The left panels are tests results for the Miller-Scalo mass 
function, and the right panels for a power law mass function. The top three 
panels (left and right) show the effects of incorrect assumptions about the 
cluster age in the de-reddening process.  Single-age assumptions give
the worst results with two effects occurring.  The first is a general shift
of the recovered mass function towards higher masses as the age assumption
is moved to older ages, due simply to the decrease in luminosity with age
for a given mass star.  The second effect is a ``kinking'' in the recovered
mass function which is caused by considerable flattening of single isochrones
in the 0.3-0.1 M$_\odot$ range compared to higher and lower masses in the \khk
diagram.  When a range of ages is assumed in the de-reddening, instead of just
a single age, this effect is smeared out.  Note that there is little
difference between the panels which assume the correct age distribution,
log-uniform between 10$^5$ and 10$^6$ yr, and the panels which assume a
somewhat broader age distribution, log-uniform between 3x10$^4$ and
3x10$^6$ yr. The bottom panels (left and right) show the effects of an
incorrect assumption about the near-infrared excess.  When no infrared excess
is allowed for in the de-reddening process too much power is given in the
recovered mass function to higher masses relative to lower masses.

To summarize our test results, we find that we can recover the input mass 
function with some reasonableness in all cases where we know the correct 
stellar age and near-infrared excess distributions, and in most cases 
where we assume somewhat (but not grossly) 
incorrect representations for these distributions. 
The worst results are obtained
when the cluster consists of a uniform age distribution, but a single age is
assumed to derive the mass function. Since the spectroscopic data for the ONC 
indicate an age spread, this is in fact the least applicable case for 
this study.  Based upon these test results, we expect that our procedure to 
recover the input mass function performs well over the mass range 
0.03$<$M/M$_\odot$$<$1, and that it can distinguish between mass functions 
that slowly fall across the hydrogen burning limit, 
as N(log M) $\propto$ M$^{0.37}$, and those that slowly
rise, as N(log M) $\propto$ M$^{-0.35}$.

\subsection{Results on the ONC Stellar/Sub-stellar Mass Function}

\begin{figure}
\epsscale{0.85}
\vskip -0.5truein
\plotone{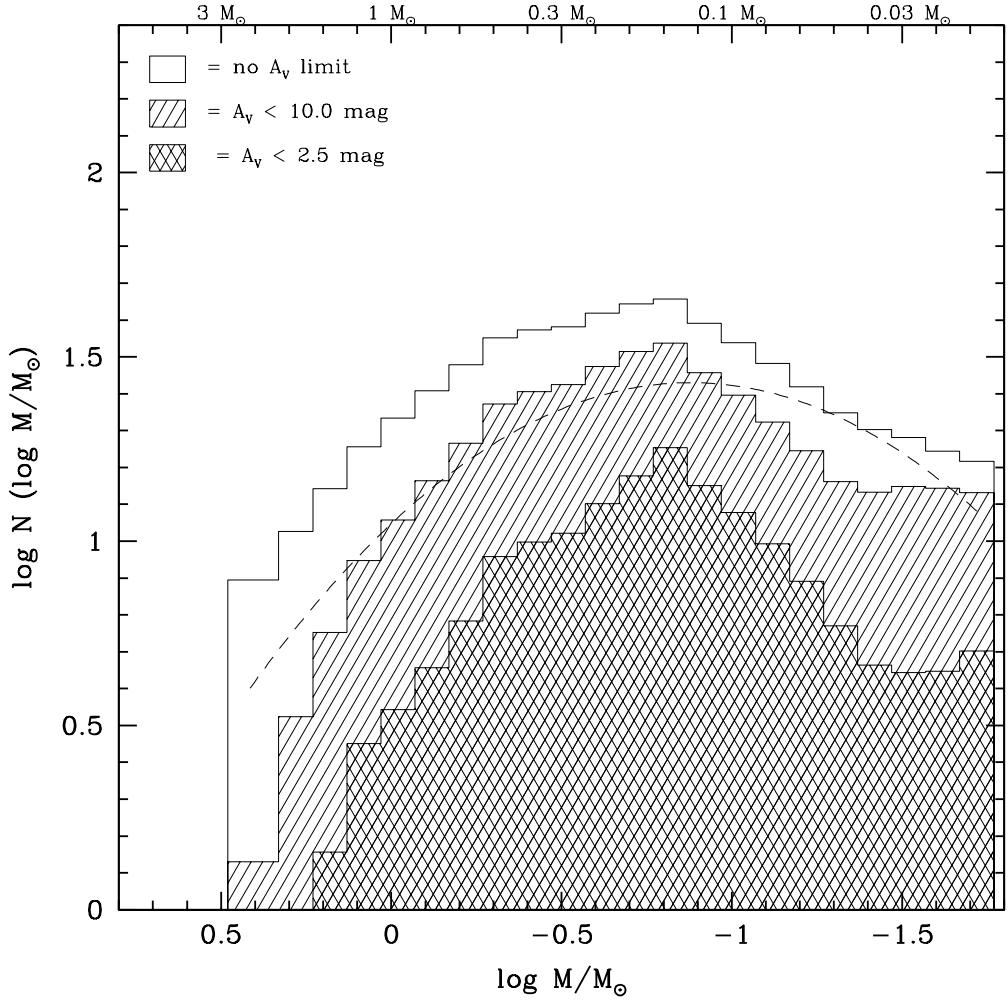}
\figcaption{
Derived ONC mass spectrum under three different extinction cuts.
The nonlinearity/saturation limit of our observations means that we are 
fully sensitive to stars with M$<$1.5 M$_\odot$ only while the full sensitivity
low-mass mass limit is M$=$0.02 M$_\odot$, for A$_V <$ 10 mag.  
A Miller-Scalo function normalized to the total number of stars in the 
A$_V <$ 10 mag distribution is shown for comparison (dashed line).  
Our data indicate that the mass function {\it in the inner ONC} declines
across the hydrogen burning limit into the brown dwarf regime, perhaps
with a somewhat narrower log-normal distribution than Miller-Scalo.
\label{fig:it}
}
\epsscale{1.00}
\end{figure}

Using the procedure we have described and tested above, we present in 
Figure~\ref{fig:it} the ONC mass function resulting from our best determinations
of the appropriate stellar age (Figure~\ref{fig:agehist}) and near-infrared 
excess (Figure~\ref{fig:irxshist}) distributions.  Of the 658 stars with
suitably good H and K photometry going in to this analysis, we recover 598 
when we integrate over this mass function.  The loss of $\sim$9\% is due to 
color-magnitude diagram locations
(spread by photometric errors; see Figure~\ref{fig:khk.data_fieldstars}a)
with no solution inside the bounds of the mass grid considered in this 
analysis given the assumed age and near-infrared excess distributions.  
Since bright massive stars 
can be detected through larger values of extinction than faint brown dwarfs, 
we also plot the mass function for only those objects meeting certain extinction
criteria: first, only those with A$_V < 10$ mag, the highest extinction
level to which 0.02 M$_\odot$ objects can be detected given the sensitivity 
limits of our survey, and second, only those with A$_V < 2.5$ mag,
the extinction limit to which 0.1 M$_\odot$ objects could be detected in 
the optical spectroscopic survey by Hillenbrand (1997), to which we compare 
our infrared photometric results below. Of the total of 598 sources 
in the mass function of Figure~\ref{fig:it} (open histogram), 
67\% have A$_V < 10$ mag while 28\% have A$_V < 2.5$ mag.

As shown in Figure~\ref{fig:it}, the stellar/sub-stellar mass function 
in the ONC peaks near $\sim$ 0.15 M$_\odot$ and is clearly falling 
across the hydrogen burning limit into the brown dwarf regime -- 
regardless of the adopted extinction limit, which affects the shape of
the mass function only at the higher masses.  We have investigated 
the robustness of Figure~\ref{fig:it} for different plausible age ranges 
(e.g. log-uniform between 3x10$^4$ and 3x10$^6$ yr instead of  
between 1x10$^5$ and 1x10$^6$ yr), with and without a near-infrared excess 
distribution, and also with and without subtraction of field stars.  The same 
basic conclusion is found.  A power law fit to the declining inner ONC 
mass function for A$_V <$ 10 mag between 0.03 M$_\odot$ and 0.2 M$_\odot$ 
has a slope of 0.57 $\pm$ 0.05 (in logarithmic units), 
where the uncertainties reflect only the residuals of the 
least squares fit to the data. Our best determination of the inner ONC 
mass function is inconsistent at the $>10\sigma$ level with a mass function 
that is flat or rising across the hydrogen burning limit.


According to the tests of our methodology (section 6.3.2), there are two ways
to add power at low masses relative to higher masses and thus
produce a less steeply declining or even flat slope across
the hydrogen-burning limit: by making the cluster age much
younger than we have assumed, and/or by making the near-infrared
excesses much larger than we have assumed. We find neither of these options 
probable given the characteristics of the optically visible stars in the region,
and hence we conclude that the inner ONC mass function is indeed declining.  
We have shown the accuracy to which our methodology recovers a known input mass function 
in Figure~\ref{fig:testmodels}.  Based on fits over the same 0.03-0.2 M$_\odot$ mass range
we consider for the data, we conclude that our method recovers the correct slope
of the input mass function to within $<$0.05. Combining this methodology error with the
r.m.s. fitting error of $\pm$0.05 discussed above, we estimate the total error 
on the slope derived here for the ONC mass function across the hydrogen burning limit at $<$0.1.
We offer the following two additional cautions to any interpreters of our results.  

First, we emphasize that the detailed shape of the mass function derived from data 
is still subject to dependence on theoretical tracks and isochrones 
(D'Antona \& Mazzitelli; 1997, 1998 in this case), and on the calibrations 
used in converting between effective temperature / luminosity and 
\khk color/magnitude (discussed in section 6.1.3).  

\begin{figure}
\vskip -0.5truein
\plotone{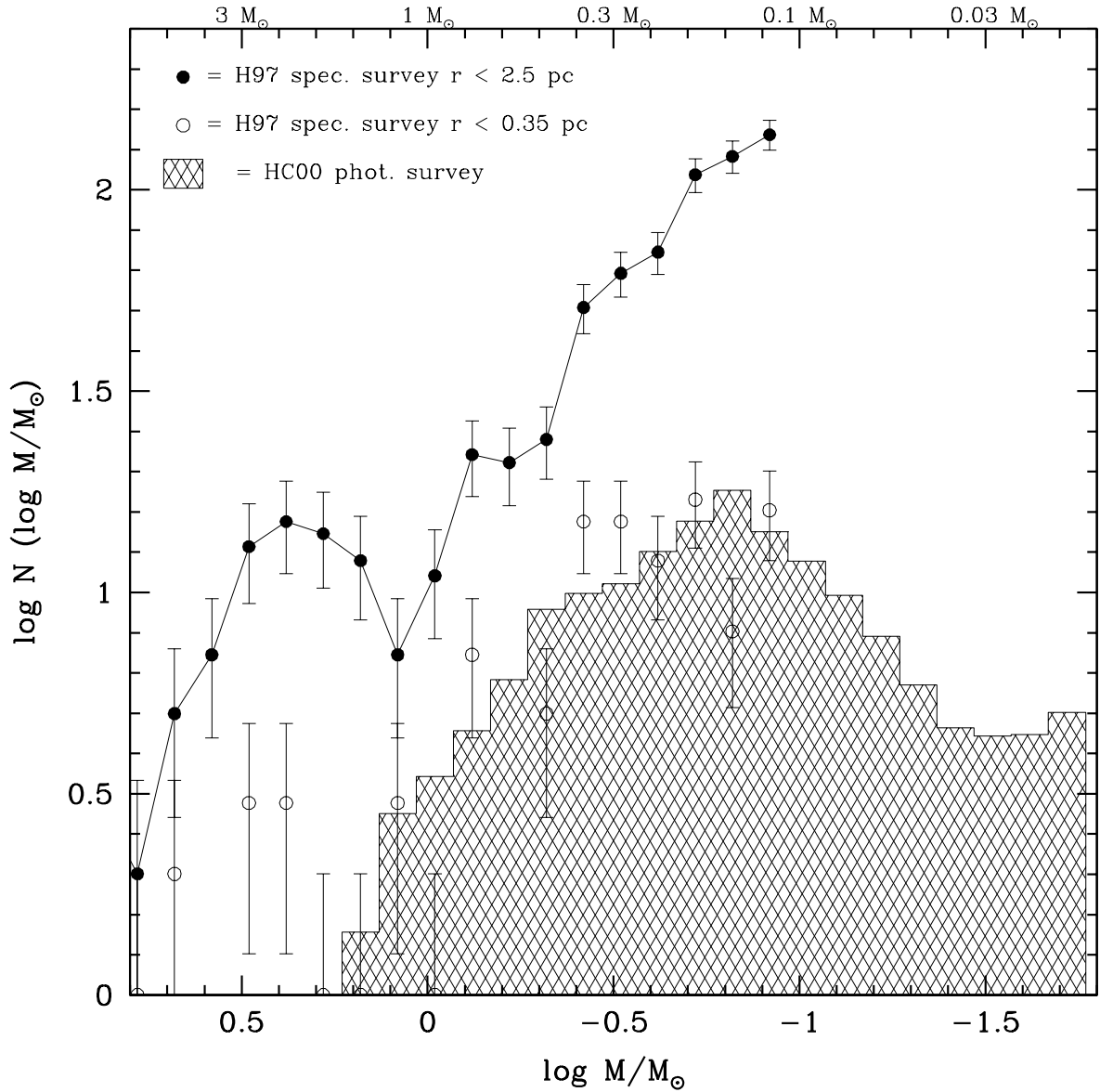}
\figcaption{Comparison of the ONC mass spectrum derived from optical 
spectroscopic techniques with that derived here using infrared photometric
techniques.  Filled circles are the same spectroscopic data as in the top panel
of Figure~\ref{fig:optical.newvsold}, now limited to 
A$_V<$ 2.5 mag leaving 758 stars.  Open circles represent that portion of
the spectroscopic data located within the same spatial area as our NIRC
data, also limited to A$_V<$ 2.5 mag leaving 120 stars.  Histogram is the NIRC 
mass function for extinction A$_V <$ 2.5 mag.  No normalization has been
applied to these curves.  Note the general agreement
between the optical spectroscopic results and the near-infrared photometric
results in the mass completeness and the spatial area regimes where they
overlap (open circles vs hatched histogram).  Note also the disagreement 
between the shape of the mass spectrum derived for the inner ONC (r$<$0.35 pc;
open circles) vs the greater ONC (r$<$2.5 pc; filled circles).
\label{fig:compare.to.optical}
}
\end{figure}

Second, we emphasize that   our derived mass function is valid only 
for the inner 0.71 pc $\times$ 0.71 pc of the ONC cluster, which extends 
at least 8-10 pc in length and 3-5 pc in width. 
Our conclusions may not apply to the ONC as a whole where some evidence 
for general mass segregation has been found by Hillenbrand (1997) and 
Hillenbrand \& Hartmann (1998).  In Figure~\ref{fig:compare.to.optical}
we compare the mass function derived here for the inner cluster using
near-infrared photometry to that derived previously by us using optical 
photometry and spectroscopy.   
The histogram is the mass function of Figure~\ref{fig:it} with an extinction
limit of A$_V<$ 2.5 mag, for consistency with the effective extinction limit 
of the optical data.  Solid symbols represent the full dataset from  
Hillenbrand (1997) while open symbols represent only that portion 
of the data which are spatially coincident with the near-infrared photometry 
used to derive the histogram (i.e. the inner 0.35 pc or so).  The same
A$_V<$ 2.5 mag imposed on the infrared data has also been imposed on 
the optical data.  As noted
in reference to Figure~\ref{fig:optical.newvsold}, the updated pre-main
sequence tracks and the updated transformations between observational 
and theoretical quantities adopted in this paper have caused a shift 
in our interpretation of the data presented by Hillenbrand (1997).   
The large-scale ONC mass function (solid symbols) now appears to be rising 
to the limit of that survey.  The inner ONC mass function (open symbols),
however, appears to flatten below $\sim$0.3 M$_\odot$.  This flattening
is confirmed by the near-infrared photometric analysis presented here,
and in fact is the beginning of a turnover in the mass function above
the hydrogen burning limit and extending down to at least 30 M$_{Jupiter}$.

\section{Discussion}

Our analysis of the mass distribution in the inner ONC agrees with that 
of McCaughrean et al. (1995) in that there is ``a substantial but not dominant 
population of young hot brown dwarfs'' in the inner ONC.  Although we do find
$\sim$80 objects with masses in the range 0.02-0.08 M$_\odot$, the overall
distribution of masses is inconsistent with a mass function that rises 
across the stellar/sub-stellar boundary.  Instead, we find that the most likely 
form of the mass function in the inner ONC is one that peaks around 
0.15 M$_\odot$ and then declines across the hydrogen-burning limit to the 
mass limit of our survey, 0.02 M$_\odot$.  The best-fit power-law for the
decline, N(log M) $\propto$ M$^{0.57}$, is steeper than that predicted 
by the log-normal representation of the Miller-Scalo initial mass function, 
N(log M) $\propto$ M$^{0.37}$ if forced to a power-law (see Figure~\ref{fig:it}).

How do our results compare to other determinations of the 
sub-stellar mass function? Thusfar there have been few actual measurements 
of the sub-stellar mass function which are not either lower limits or dominated
by  incompleteness corrections or small-number statistics.  
We can compare our results for the inner ONC only to those in the Pleiades 
(Bouvier et al. 1998; Festin 1998)
and the solar neighborhood (Reid et al. 1999), and we find some 
differences.  Converting the logarithmic units used thusfar in this paper
(N(log M) $\propto$ M$^\Gamma$)
to the linear units adopted by others (N(M) $\propto$ M$^\alpha$)  
we find a mass function slope across the hydrogen-burning limit of
$\alpha = \Gamma - 1$ = -0.43.  In the Pleiades, Bouvier et al. 
find $\alpha$ = -0.6 while Festin finds $\alpha$ in the range 0 to -1.0. 
In the solar neighborhood, Reid et al. find $\alpha$ in the range -1.0 to -2.0 
with some preference for the former value.
The methods used by these different authors for arriving at the slope of
the mass function are very different, thus rendering somewhat difficult
any interpretation of the comparison.  Furthermore, it is not clear that the
mass function in the center of a dense and violent star-forming environment
should bear any similarity to the mass function in a lower-density,
quiescent older cluster, or that either of these cluster mass functions
should look anything like the well-mixed, much older local field star 
population.
Nevertheless, if comparisons can be made, the inner ONC seems to have 
a shallower slope than that found in any other region where measurements
have been made; recall as well that the inner ONC mass function 
appears shallower than the overall ONC mass function 
(see Figure~\ref{fig:compare.to.optical}).

\section{Conclusions}

We have introduced a new method for constraining the stellar/sub-stellar
mass distribution for optically invisible stars in a star-forming region.  
A comparative review of the various techniques already in use for measuring 
mass functions in star-forming regions is presented by Meyer et al. (2000).
These techniques range from studies of observed K-magnitude histograms 
(e.g. Muench et al. 2000), to discrete de-reddening of infrared color-magnitude
diagrams (e.g. Comeron, Rieke, \& Rieke 1996), 
to the assembly of photometric and spectroscopic 
data from which HR diagrams are created (e.g. Luhman \& Rieke 1998).
Our method is a variation on and an improvement to the discrete de-reddening
of color-magnitude diagrams since we fully account for distributions in the
relevant parameters instead of assuming a mean value for them.  However,
our method is not as good as a complete photometric - plus - spectroscopic
survey since we produce only a mass probability distribution for each star,
not a uniquely determined mass.  Nonetheless, we believe that the statistical 
nature of our method does provide the most rigorously established constraint 
to date from photometry alone on the stellar mass function in a star-forming region.

We have used information from previous studies of optically visible stars 
in the ONC to derive plausible functional forms for the stellar age and 
the circumstellar near-infrared excess distributions
in the innermost regions studied here. We assume that these distributions 
apply equally well to the optically invisible population.  
We find a mass function for the inner 0.71 pc x 0.71 pc of the ONC which rises 
to a peak around 0.15 M$_\odot$ and then declines
across the stellar/sub-stellar boundary as N(log M) $\propto$ M$^\Gamma$
with slope $\Gamma=0.57$.  This measurement is of the primary 
star/sub-star mass function only, and should be adjusted by the (currently
unknown) companion mass function in order to derive the 
``single star mass function,'' if desired.

We find strong evidence that the 
shape of the mass function for this inner ONC region is different from that
characterizing the ONC as a whole, in the sense that the flattening and
turning over of the mass function occurs at higher mass in the inner region
than in the overall ONC.  In fact, the shape of mass function for the overall 
ONC is currently unconstrained across the stellar/sub-stellar boundary, 
and appears now based on the most recent theoretical tracks and conversions
between the theory and observables used in this paper,
to continue to rise to at least 0.12 M$_\odot$.

\acknowledgements
We thank Mike Liu and James Graham for sharing their method and source code
for ``de-bleeding'' of NIRC images.  We thank Andrea Ghez for providing 
her image distortion coefficients.  We thank Keith Matthews for consultation 
regarding these and other NIRC features. Shri Kulkarni and Ben Oppenheimer
suggested a mutually advantageous exchange of telescope time which
enabled us to obtain the Z-band observations.  Ted Bergin kindly provided 
his C$^{18}$O data to us and Richard Wainscoat gave us a base code for his
star count model.  This publication makes use of data products from 
the Two Micron All Sky Survey, which is a joint project of the 
University of Massachusetts and the Infrared Processing and Analysis
Center, funded by the National Aeronautics and Space Administration 
and the National Science Foundation. 
LAH acknowledges support from NASA Origins of Solar Systems 
grant \#NAG5-7501.  JMC acknowledges support from NASA Long Term Space 
Astrophysics grant \#NAG5-8217, and from the Owens Valley Radio Observatory 
which is operated by the California Institute of Technology through NSF 
grant \#96-13717.

\appendix
\section{The Infrared Variable 2MASSJ053448-050900 = AD95-1961}

\begin{figure}
\vskip -0.5truein
\plotone{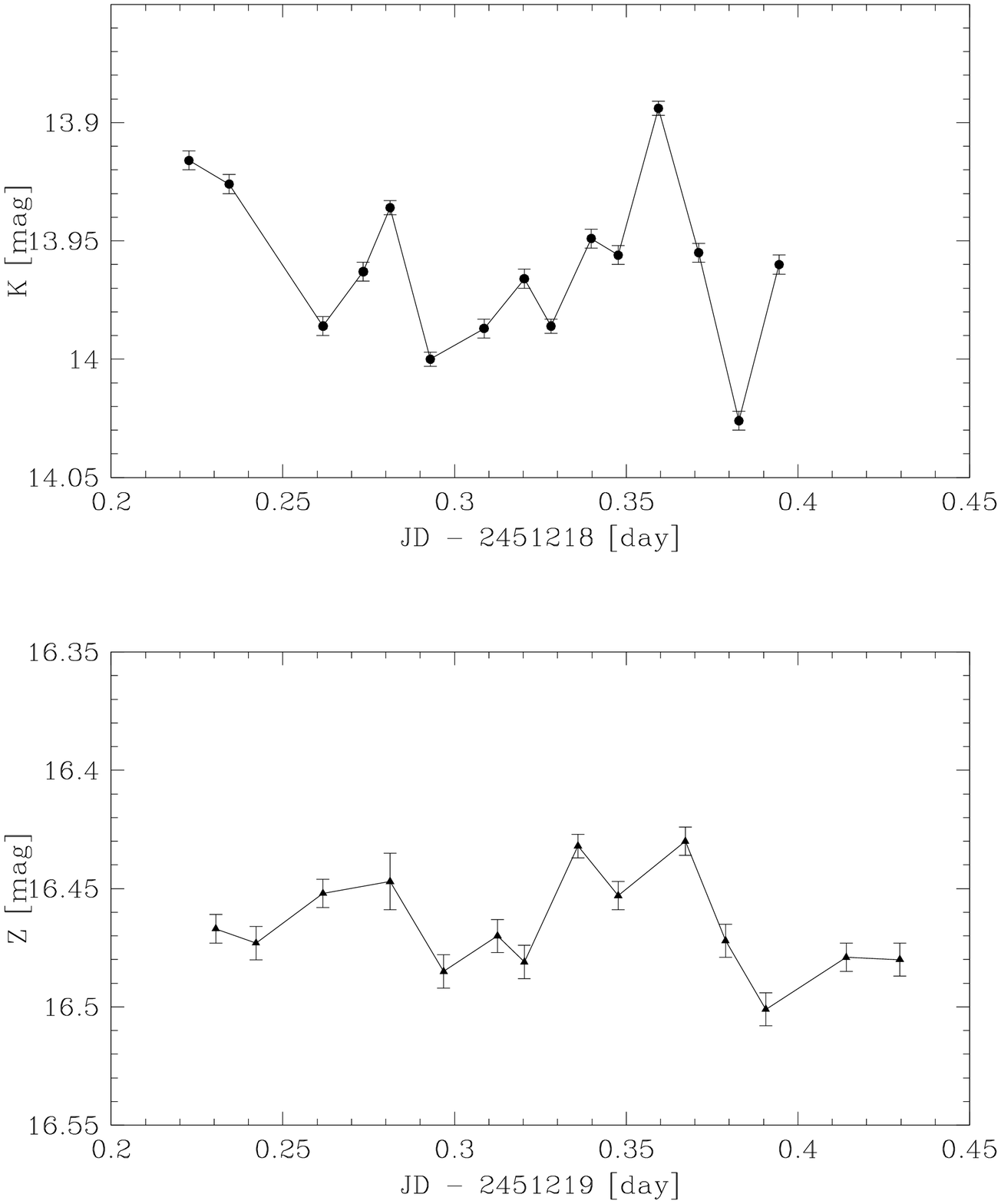}
\vskip -0.5truein
\figcaption{Infrared variable star 2MASSJ053448-050900 = AD95-1961.
This object is located approximately 15' northeast of our mosaic center
and was observed as a local standard for the purpose of atmospheric extinction 
calibration.
The observations plotted were taken 12-15 minutes apart and show variations at
the 0.05-0.1 mag level.  Similar variability on similar timescales may be 
a common feature of the young stellar objects the ONC.
\label{fig:variablestar}
}
\end{figure}

In this appendix we report on short timescale variability at infrared wavelengths
of a star located approximately 15' north-west of the ONC.  
Our observing procedure in constructing our 5.1' x 5.1' mosaic
with NIRC was to scan across a row at constant declination, then move off
to a sky position and obtain five measurements of the sky which were then
averaged and subtracted from each frame in our ONC mosaic.  We intentionally
chose a sky field which included a relatively bright (K$\approx$ 14.0 mag)
star in order to monitor the atmospheric extinction as part of
our normal data acquisition.  However, while our set of absolute
standards from Persson et al. (1998) matched nominal NIRC zero points 
and nominal Mauna Kea extinction curves with airmass, our local standard
exhibited significant flux variations (Figure~\ref{fig:variablestar}). 
The amplitude of the variations is about 0.1 mag, and the timescale is less 
than the separation of our observations, about 10-12 minutes.  

2MASSJ053448-050900 is also catalogued as AD95-1961 (Ali \& DePoy 1995).  
This star has infrared fluxes of K=14.03 mag, H=14.43 mag, J=15.46 mag
from the 2MASS survey and optical fluxes of I$\approx$17.5 mag, 
V$\approx$21.1 mag from our own unpublished CCD observations.
These colors are consistent with those of low-mass ONC proper motion members.

The short-term photometric behavior of this relatively isolated and otherwise
nondescript star located in the outer regions of the ONC
may in fact be a general feature of all young stellar objects.
Infrared monitoring studies of young clusters are needed in order to
quantify the nature and constrain the causes of this variability.

\clearpage

\clearpage

\begin{tiny}
\begin{deluxetable}{r r r r r r r r}
\tablenum{1}
\tablewidth{0pt}
\tablecolumns{4}
\tablecaption{Keck/NIRC Photometry and Astrometry for the Inner ONC}
\tablehead{
\colhead{ID} &
\colhead{Position} &
\colhead{K} &
\colhead{H} &
\colhead{$\sigma$(K)} &
\colhead{$\sigma$(H)} &
\colhead{Optical ID} &
\colhead{Alternate Infrared ID}
\\[.2ex]
\colhead{} &
\colhead{[J2000.]} &
\colhead{[mag]} &
\colhead{[mag]} &
\colhead{[mag]} &
\colhead{[mag]} &
\colhead{} &
\colhead{}
}
\startdata

\enddata
\tablenotetext{}{
ID = running number (note that line 187 is missing as this source is no longer 
believed real);
Position = J2000. coordinates referenced to the ACT catalog;
K, H = photometry measured with Keck/NIRC;
$\sigma$(K), $\sigma$(H) = IRAF errors in photometry; 
Optical ID = identification numbers as listed in Hillenbrand (1997)
or from O'Dell \& Wong (1996);
Alternate Infrared ID = identification numbers as listed in
McCaughrean \& Stauffer (1994), Lonsdale et al. (1982),
Downes et al. (1981), or Rieke, Low, and Kleinmann (1973).}
\label{tab:data}
\end{deluxetable}

[THIS TABLE CAN BE FOUND IN ascii FORMAT AT
{\tt http://astro.caltech.edu/$\sim$lah/papers.html}
]
\end{tiny}

\begin{small}
\begin{deluxetable}{c r r r}
\tablenum{2}
\tablewidth{0pt}
\tablecolumns{4}
\tablecaption{Internal errors in artifical star photometry}
\tablehead{
\colhead{brightness range} &
\colhead{$<$ 0.02 mag} &
\colhead{$<$ 0.05 mag} &
\colhead{$<$ 0.10 mag}
\\[.2ex]
\colhead{[mag]} &
\colhead{[\%]} &
\colhead{[\%]} &
\colhead{[\%]} 
}
\startdata

K=15.00-15.25 &     69&   89&   96\nl
K=15.25-15.50 &     66&   85&   93\nl
K=15.50-15.75 &     60&   84&   91\nl
K=15.75-16.00 &     52&   81&   91\nl
K=16.00-16.25 &     44&   77&   87\nl
K=16.25-16.50 &     31&   74&   85\nl
K=16.50-16.75 &     15&   68&   85\nl
K=16.75-17.00 &      4&   59&   81\nl
K=17.00-17.25 &      1&   49&   74\nl
K=17.25-17.50 &      0&   36&   72\nl
K=17.50-17.75 &      0&   19&   59\nl
K=17.75-18.00 &      0&    8&   48\nl
K=18.00-18.25 &      0&    1&   38\nl
K=18.25-18.50 &      0&    1&   16\nl
         &         &       &       \nl
H=15.00-15.25 &     75&   88&   96\nl
H=15.25-15.50 &     69&   88&   94\nl
H=15.50-15.75 &     63&   85&   92\nl
H=15.75-16.00 &     56&   82&   91\nl
H=16.00-16.25 &     39&   78&   89\nl
H=16.25-16.50 &     32&   68&   85\nl
H=16.50-16.75 &     24&   61&   83\nl
H=16.75-17.00 &     15&   52&   77\nl
H=17.00-17.25 &      7&   42&   72\nl
H=17.25-17.50 &      3&   34&   64\nl
H=17.50-17.75 &      1&   27&   58\nl
H=17.75-18.00 &      1&   16&   48\nl
H=18.00-18.25 &      1&    8&   42\nl
H=18.25-18.50 &      1&    4&   36\nl

\enddata
\label{tab:internalerrs}
\end{deluxetable}
\end{small}

\begin{small}
\begin{deluxetable}{r r r r}
\tablenum{3}
\tablewidth{0pt}
\tablecolumns{4}
\tablecaption{Offset between input and recovered magnitudes}
\tablehead{
\colhead{brightness range} &
\colhead{0.0-0.2 pc } &
\colhead{0.2-0.4 pc } &
\colhead{0.4-0.5 pc } 
\\[.2ex]
\colhead{[mag]} &
\colhead{[mag] } &
\colhead{[mag] } &
\colhead{[mag] } 
}
\startdata

K=15.00-15.25 & 0.01$\pm$ 0.01 &  0.00$\pm$ 0.01 &  0.00$\pm$ 0.00 \nl
K=15.25-15.50 & 0.00$\pm$ 0.02 &  0.00$\pm$ 0.01 &  0.00$\pm$ 0.01 \nl
K=15.50-15.75 & 0.00$\pm$ 0.02 &  0.00$\pm$ 0.01 &  0.00$\pm$ 0.01 \nl
K=15.75-16.00 & 0.01$\pm$ 0.04 &  0.00$\pm$ 0.01 &  0.00$\pm$ 0.00 \nl
K=16.00-16.25 & 0.02$\pm$ 0.04 &  0.00$\pm$ 0.01 &  0.01$\pm$ 0.02 \nl
K=16.25-16.50 & 0.05$\pm$ 0.04 &  0.00$\pm$ 0.02 &  0.00$\pm$ 0.01 \nl
K=16.50-16.75 & 0.02$\pm$ 0.05 &  0.01$\pm$ 0.03 & -0.01$\pm$ 0.02 \nl
K=16.75-17.00 & 0.04$\pm$ 0.06 &  0.01$\pm$ 0.03 & -0.01$\pm$ 0.02 \nl
K=17.00-17.25 & 0.03$\pm$ 0.08 &  0.02$\pm$ 0.03 &  0.01$\pm$ 0.03 \nl
K=17.25-17.50 & 0.07$\pm$ 0.11 &  0.01$\pm$ 0.04 & -0.01$\pm$ 0.02 \nl
K=17.50-17.75 & 0.05$\pm$ 0.15 &  0.02$\pm$ 0.06 &  0.04$\pm$ 0.04 \nl
K=17.75-18.00 & 0.04$\pm$ 0.18 &  0.03$\pm$ 0.08 &  0.05$\pm$ 0.06 \nl
K=18.00-18.25 &-0.15$\pm$ 0.46 &  0.04$\pm$ 0.10 &  0.02$\pm$ 0.06 \nl
K=18.25-18.50 &-0.06$\pm$ 0.39 &  0.05$\pm$ 0.13 &  0.02$\pm$ 0.08 \nl
              &           &         &         \nl
H=15.00-15.25 & 0.01$\pm$ 0.04 &  0.00$\pm$ 0.02 &  0.00$\pm$ 0.01 \nl
H=15.25-15.50 & 0.01$\pm$ 0.05 &  0.00$\pm$ 0.02 &  0.01$\pm$ 0.01 \nl
H=15.50-15.75 & 0.01$\pm$ 0.06 &  0.00$\pm$ 0.02 &  0.01$\pm$ 0.01 \nl
H=15.75-16.00 & 0.00$\pm$ 0.09 &  0.01$\pm$ 0.03 &  0.01$\pm$ 0.01 \nl
H=16.00-16.25 & 0.02$\pm$ 0.09 &  0.02$\pm$ 0.03 &  0.00$\pm$ 0.02 \nl
H=16.25-16.50 & 0.03$\pm$ 0.11 &  0.02$\pm$ 0.04 &  0.00$\pm$ 0.02 \nl
H=16.50-16.75 & 0.03$\pm$ 0.13 &  0.03$\pm$ 0.05 &  0.00$\pm$ 0.03 \nl
H=16.75-17.00 & 0.04$\pm$ 0.19 &  0.03$\pm$ 0.07 & -0.01$\pm$ 0.03 \nl
H=17.00-17.25 & 0.02$\pm$ 0.24 &  0.03$\pm$ 0.07 & -0.01$\pm$ 0.04 \nl
H=17.25-17.50 &-0.02$\pm$ 0.33 &  0.05$\pm$ 0.10 & -0.01$\pm$ 0.06 \nl
H=17.50-17.75 &-0.38$\pm$ 0.58 &  0.03$\pm$ 0.12 & -0.03$\pm$ 0.07 \nl
H=17.75-18.00 &-0.63$\pm$ 0.63 &  0.02$\pm$ 0.14 & -0.04$\pm$ 0.09 \nl
H=18.00-18.25 &-0.87$\pm$ 0.69 &  0.00$\pm$ 0.22 & -0.05$\pm$ 0.11 \nl
H=18.25-18.50 &-1.12$\pm$ 0.71 & -0.04$\pm$ 0.29 & -0.06$\pm$ 0.13 \nl

\enddata
\label{tab:externalerrs}
\end{deluxetable}
\end{small}

%
%

\clearpage

\end{document}